\documentclass[aps,prb,reprint,superscriptaddress]{revtex4-1}

\usepackage[utf8x]{inputenc} 
\usepackage{amsmath}
\usepackage[dvips]{graphicx}
\usepackage{amssymb}
\usepackage{color}
\usepackage{MnSymbol}
\usepackage{mathtools}

\renewcommand{\d}{\mathrm{d}}

\newcommand{\e}{\mathrm{e}}
\DeclareFontEncoding{FML}{}{}%
\DeclareFontSubstitution{FML}{futm}{m}{it}%
\DeclareSymbolFont{fourier}{FML}{futm}{m}{it}%
\DeclareMathSymbol{\partialup}{\mathord}{fourier}{130} 
\usepackage{upgreek}
\renewcommand{\pi}{\uppi}
\newcommand{\ddt}{\frac{\partialup}{\partialup t}}
\renewcommand{\delta}{\updelta}
\renewcommand{\i}{\mathrm{i}}
\newcommand{\tr}{\mathrm{tr}}
\newcommand{\Oenv}{\varOmega_\mathrm{env}}
\newcommand{\env}{\mathrm{env}}
\renewcommand{\Im}{\mathrm{Im}}
\renewcommand{\Re}{\mathrm{Re}}
\newcommand{\wC}{\omega_\mathrm{C}}
\newcommand{\wL}{\omega_\mathrm{L}}
\newcommand{\eins}{\llangle b^\dagger b\rrangle}
\newcommand{\zweiL}{\llangle P_{3,2} b^\dagger\rrangle}
\newcommand{\zweiV}{\llangle P_{2,1} b^\dagger\rrangle}
\newcommand{\dreienv}{\llangle P_{3,1} b^\dagger\rrangle_\env}
\newcommand{\drei}{\llangle P_{3,1} b^\dagger\rrangle}
\newcommand{\vierLenv}{\langle P_{2,1}\rangle_\env}
\newcommand{\vierVenv}{\langle P_{3,2}\rangle_\env}
\newcommand{\fuenf}{\langle P_{1,1} \rangle}
\newcommand{\sechs}{\langle P_{3,3} \rangle}
\newcommand{\sieben}{\llangle \sigma_{2,1} b^\dagger b\rrangle}
\newcommand{\acht}{\llangle \sigma_{3,2} b^\dagger b\rrangle}
\newcommand{\neunLenv}{\llangle P_{2,1} b^\dagger b\rrangle_\env}
\newcommand{\neunVenv}{\llangle P_{3,2} b^\dagger b\rrangle_\env}
\newcommand{\DL}{\varDelta_\mathrm L}
\newcommand{\DC}{\varDelta_\mathrm C}
\newcommand{\V}{\mathrm V}
\newcommand{\R}{\mathrm R}
\newcommand{\opt}{\mathrm{opt}}

\begin{document}


\title{Microscopic theory of cavity-enhanced  single-photon emission from optical two-photon Raman processes}

\author{Dominik Breddermann}
\affiliation{Department of Physics and Center for Optoelectronics and Photonics Paderborn (CeOPP), Paderborn University, Warburger Strasse 100, 33098 Paderborn, Germany}
\author{Tom Praschan}
\affiliation{Department of Physics and Center for Optoelectronics and Photonics Paderborn (CeOPP), Paderborn University, Warburger Strasse 100, 33098 Paderborn, Germany}
\author{Dirk Heinze}
\affiliation{Department of Physics and Center for Optoelectronics and Photonics Paderborn (CeOPP), Paderborn University, Warburger Strasse 100, 33098 Paderborn, Germany}
\author{Rolf Binder}
\affiliation{College of Optical Sciences, University of Arizona, Tucson, Arizona 85721, USA}
\author{Stefan Schumacher}
\email{stefan.schumacher@upb.de}
\affiliation{Department of Physics and Center for Optoelectronics and Photonics Paderborn (CeOPP), Paderborn University, Warburger Strasse 100, 33098 Paderborn, Germany}
\affiliation{College of Optical Sciences, University of Arizona, Tucson, Arizona 85721, USA}

\date{\today}

\begin{abstract}
We consider cavity-enhanced single-photon generation from stimulated two-photon Raman processes in three-level systems. We compare four fundamental system configurations, one $\Lambda$-, one V- and two ladder ($\Xi$-) configurations. These  can be realized as subsystems of a single quantum dot  or of quantum-dot molecules. For a new microscopic understanding of the Raman process, we analyze the Heisenberg equation of motion applying the cluster-expansion scheme. Within this formalism an exact and rigorous definition of a cavity-enhanced Raman photon via its corresponding Raman correlation is possible. This definition for example enables us to systematically investigate the on-demand potential of Raman-transition-based single-photon sources. The four system arrangements can be divided into two subclasses, $\Lambda$-type and V-type, which exhibit strongly different Raman-emission characteristics and Raman-emission probabilities.  Moreover, our approach reveals whether the Raman path generates a single photon or just induces destructive quantum interference with other  excitation paths. Based on our findings and as a first application, we gain a more detailed understanding of experimental data from the literature.  Our analysis and results are also transferable to the case of atomic three-level-resonator systems, and can be extended to more complicated multi-level schemes.
\end{abstract}

\pacs{}

\maketitle

\section{Introduction}

\begin{figure}[t!]
 \centering
 \includegraphics[width=0.47\textwidth]{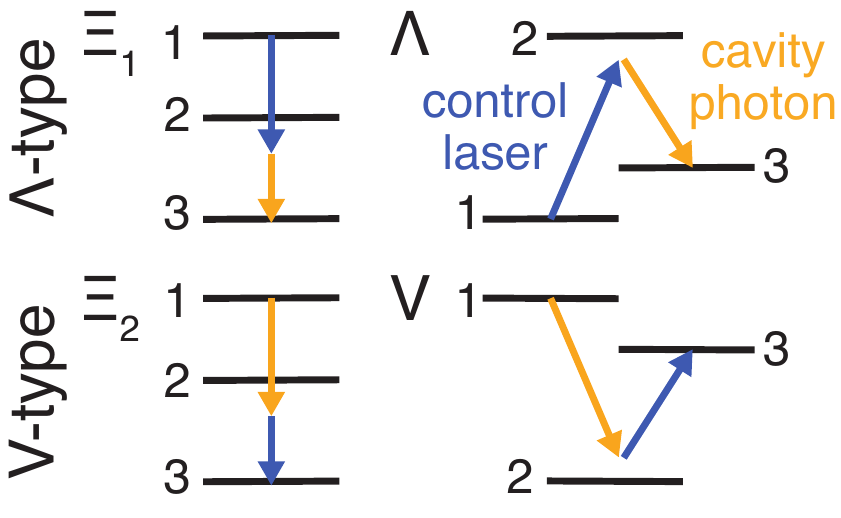}
 \caption{Three-level systems in photon-emission configuration. Shown is the two-photon Raman process where one photon is driven by an external laser field (blue arrow) and the other photon (yellow arrow) is emitted spontaneously into a cavity mode. We label the electronic configurations 1 (initial state), 2 (intermediate state) and 3 (final state), according to the direction of the respective two-photon path. The transitions between energy levels 1 and 2 and between energy levels 2 and 3 are assumed to have orthogonal dipole moments such that they can be excited selectively. The systems are labeled $\Xi_1$, $\Lambda$, $\Xi_2$ and V and can be grouped into two subclasses, $\Lambda$-type and V-type. In a quantum dot, the ladder configuration can be realized via biexciton state (1), exciton state (2) and ground state (3). The $\Lambda$-configuration is commonly realized in a charged quantum dot or a quantum-dot molecule in the presence of a magnetic field. In the V-configuration, states 1 and 3 could be the finestructure-split excitons and state 2 the ground state. Atomic realizations of the four systems are abundant.}
\label{fig:systems}
\end{figure}

Light emission from laser-enhanced two-photon processes, also referred to as optical two-photon Raman processes, has been extensively studied in atomic three-level systems \cite{yatsiv1968enhanced,braunlich1970detection,yoo1985dynamical,hennrich2000vacuum}, including quantum-network \cite{cirac1997quantum} and single-photon storage applications \cite{gorshkov2007photon}. In addition to that, coherent quantum-interference phenomena mediated by related  two-photon excitations in generic three-level systems like electromagnetically induced transparency (EIT) have been thoroughly investigated\cite{boller1991observation,fleischhauer2000dark,eisaman2005electromagnetically,sen2015comparison}. Nowadays, optical Raman processes again attract attention in the context of cavity-enhanced single-photon generation from quantum-dot (QD) systems in different electronic configurations \cite{atature2006quantum,santori2009indistinguishability,sweeney2014cavity,vora2015spin,heinze2015quantum}. Moreover, an integrated nanophotonics platform \cite{sipahigil2016integrated} and single microwave photon detectors \cite{kyriienko2016continuous,inomata2016single} have recently been realized operating on Raman processes.  The setup to be considered  is sketched in Fig.~\ref{fig:systems}. Here, one photon is driven by an external control laser and the other photon is emitted spontaneously into a cavity mode. This mechanism allows for direct all-optical tunability of emission time, frequency, linewidth, spectral shape and polarization state of the emitted single photon \cite{he2013indistinguishable,heinze2015quantum,breddermann2016tailoring,beguin2017demand,pursley2018picosecond}. 

Despite all this previous work on the optical properties of the emitted light from cavity-enhanced optical Raman processes, thus far on-demand emission probability of a single photon has not been realized or studied in detail yet, neither from atoms \cite{hennrich2000vacuum,keller2004continuous} nor from a QD \cite{he2013indistinguishable}. For quantum-communication networks a minimum-extraction efficiency of $67\%$ is needed\cite{varnava2008good,he2013indistinguishable,ding2016demand}. Moreover,  emission spectra measured in recent experiments raised an open question concerning the contradictory behaviour of peak-intensity and broadening for the resonant Raman process, as discussed in the Supplementary Information of Ref.~\onlinecite{vora2015spin}. For that reason, a more detailed microscopic understanding of the Raman emission, revealing full information on the excitation processes and paths contributing to or diminishing the emission, is needed. 

In this article, we present a comprehensive microscopic analysis of cavity-enhanced single-photon emission from two-photon Raman processes. Here, we focus on three-level systems in all possible configurations where the cavity enhances the emission of one photon of the respective two-photon Raman path. This yields one  $\Lambda$-, one V- and two ladder ($\Xi$-) configurations (see Fig.~\ref{fig:systems}). We assume that cavity and control laser have orthogonal polarizations. These three-level systems -- and superpositions of them \cite{heinze2015quantum,vora2015spin} -- appear as fundamental building blocks of Raman-emission-based single-photon sources realized by single  QDs or QD molecules embedded in microresonators.  The four system configurations can be divided into two subclasses where the cavity either couples to the first state 1 (V-type) or to the final state 3 ($\Lambda$-type).  We will use the nomenclature \emph{T-type system} ($T=\Lambda$ or V) whenever talking about the whole system \emph{class}, and the nomenclature \emph{configuration} whenever  addressing a chosen \emph{representative} ($\Xi_1$, $\Xi_2$, $\Lambda$ or $\V$) of the respective class. We use the Heisenberg equation of motion together with the cluster-expansion method \cite{kira2006many} to study the single-photon emission dynamics in terms of fundamental physical processes present in the four QD-cavity systems.

\begin{figure}[t!]
 \centering
 \includegraphics[width=0.47\textwidth]{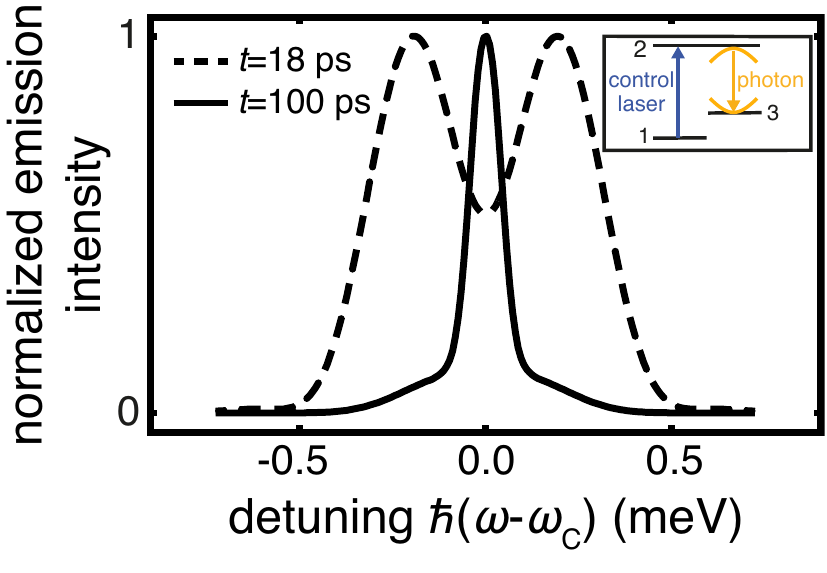}
 \caption{Time-resolved emission spectrum for resonant excitation. We show the calculated time-resolved emission spectrum\cite{eberly1977time,breddermann2016tailoring} $\mathcal S(t,\omega)=\mathrm{Re}\int_0^t\d t^\prime\int_0^{t-t^\prime}\d\tau\e^{-i\omega\tau}G^{(1)}(t^\prime,\tau)$ around the cavity frequency $\wC$ for a $\Lambda$-configuration where control laser and cavity both are resonant with the respective transition (see inset). Here, $G^{(1)}$ denotes the photon-autocorrelation function. The system is excited by a $5\uppi$ rectangular control pulse with a duration of 18 ps. Shown are two representative snapshots, one on the time scale of the Raman process (dashed line) and one on a much larger timescale (solid line). Since both spectra strongly differ in magnitude (approximately by a factor of 30), they  are normalized to their respective maximum.  Explicitly, the cavity energy is $\hbar\wC=1.296$ eV and the cavity linewidth $\hbar\kappa=186$ $\upmu$eV corresponding to a cavity quality of $Q=\wC/\kappa\approx7000$. Furthermore, a cavity-QD coupling of  $g=57$ $\upmu$eV and a QD linewidth of $\hbar\varGamma=30$ $\upmu$eV is chosen. These parameters are compatible with experimental data from Ref.~\onlinecite{vora2015spin}. Note that the spectra correspond to the excitation scenario discussed later on in Figs. \ref{fig:probdldc}d-\ref{fig:probdldc}f.}
\label{fig:interfer}
\end{figure}

This approach leads to new microscopic insight into the Raman process which, in general, cannot be extracted directly from the observation of optical spectra. As an example, Fig. \ref{fig:interfer} shows the time development of the cavity-emission spectrum for a $\Lambda$-configuration with a resonant control pulse and a resonant cavity (see also inset of the figure). On the time scale of the coherent emission (dashed line), the spectrum exhibits a pronounced dip in the center of the emission peak which is the fingerprint of destructive quantum interference of different excitation paths\cite{zhu1995quantum}.  This effect is well known from three-band or three-level excitations of zero- and higher-dimensional semiconductor structures\cite{muller2004pulse,dynes2005ac,brunner2009coherent,schubert2014sub}.  In particular, EIT \cite{harris1997electromagnetically,scully1997quantum} is a prominent example of quantum interference due to a two-photon excitation leading to a dip or even a spectral gap in the absorption spectrum \cite{phillips2003electromagnetically,marcinkevivcius2008transient,barettin2009optical}. In our case, for larger measurement durations (solid line) the interference feature is not visible anymore. Here, immediately the question arrises if the observed emission can be identified as a Raman photon or not. Another question would be which process causes the destructive interference feature observed in  the spectrum. Since neither spectral nor polarization filtering will decompose the spectrum in the resonant case, these questions can only be answered by a microscopic analysis of the emission.
In this regard and in contrast to commonly used Schr\"odinger-picture descriptions, the cluster-expansion description enables a new, rigorous and unambiguous definition of a cavity-enhanced single photon stemming from the Raman process. Moreover, this approach goes beyond established effective-Hamiltonian descriptions\cite{cirac1997quantum,del2010two} which are limited to the far-off-resonant regime of the Raman process. Here, we will follow Ref.~\onlinecite{berger2014quantum} and establish a decomposition of the full cavity-photon population $N$ into contributions $N_S$ of all emission sources $S$ corresponding to different excitation (or emission) paths,
\begin{equation}
 N=\sum_S N_S.
\label{eq:decomp0}
\end{equation} 
We will investigate under which conditions the decomposition of the full emission is dominated by the Raman source $S=\mathrm R$ and compute the corresponding emission probabilities $\mathcal P_S$. Based on our definitions, we will furthermore be able to distinguish whether the Raman process indeed generates a single \emph{Raman photon} or causes destructive interference with the emission from other sources $S$. We note that this approach can easily be extended to systems with more than three levels.

\section{Three-level systems in photon-emission configuration}

Let us first introduce the Hamiltonians of the four fundamental three-level configurations which have the form $H(t)=H_0+H_\mathrm{I}(t)$ with the noninteracting part 
\begin{equation}
 H_0=\hbar\left(\sum_{i=1}^3\omega_iP_{i,i}+\wC b^\dagger b\right)
\end{equation}
 and the light-matter interaction part $H_\mathrm{I}$. Here, we have defined the electronic operators $P_{i,j}=|i\rangle\langle j|$ and the bosonic photon annihilation (creation) operators $b^{(\dagger)}$. The energies of the electronic states and the cavity are given by $\hbar\omega_i$ and $\hbar\wC$. For the respective  configuration (see Fig.~\ref{fig:systems} for the nomenclature) and in rotating-wave approximation (RWA), $H_\mathrm{I}$ is explicitly given by
\begin{align}
 \varXi_1(t)&=\varOmega^\star(t)P_{2,1}+\varOmega(t)P_{1,2}+g\left(P_{3,2}b^\dagger+P_{2,3}b\right),  \label{eq:Xi1}\\
 \varLambda(t)&=\varOmega(t)P_{2,1}+\varOmega^\star(t)P_{1,2}+g\left(P_{3,2}b^\dagger+P_{2,3}b\right), \label{eq:Lambda}\\
 \varXi_2(t)&=g\left(P_{2,1}b^\dagger+P_{1,2}b\right)+\varOmega^\star(t)P_{3,2}+\varOmega(t)P_{2,3}, \label{eq:Xi2}\\
 V(t)&=g\left(P_{2,1}b^\dagger+P_{1,2}b\right)+\varOmega(t)P_{3,2}+\varOmega^\star(t)P_{2,3}. \label{eq:V}
\end{align}
The constant $g$ denotes the photon-material coupling and $\varOmega^{(\star)}(t)$ is the Rabi energy for absorption of (or stimulation by) the coherent external light field at time $t$. It is explicitly given by $\varOmega(t)=\varOmega_\env(t)\e^{-\i\wL t}$ where $\varOmega_\env$ denotes the envelope part and $\wL$ the excitation frequency. To be able to provide a most transparent discussion of relevant physical processes, in the present work we assume that laser field and cavity mode have orthogonal polarizations. The general discussion, however, can be extended to arbitrary selection rules. The Hamiltonians Eqs. \eqref{eq:Xi1} and  \eqref{eq:Lambda} can be transformed into each other interchanging $\varOmega$ and $\varOmega^\star$. This is also valid for Eqs. \eqref{eq:Xi2} and \eqref{eq:V}. Hence, the four  configurations can be divided into two general types, $\Lambda$-type and V-type (see also Fig.~\ref{fig:systems}).

\section{Equation-of-motion approach}
Here we only give a brief overview of the equation-of-motion approach combined with the cluster-expansion scheme which is meanwhile also established in textbooks \cite{kira2011semiconductor}.
 To analyze the system dynamics, we evaluate the Heisenberg equation of motion 
  \begin{equation}
  \i\hbar\ddt \langle O\rangle=\big\langle[O,H]\big\rangle+\sum_L\big\langle\mathcal L_L(O)\big\rangle
\label{eq:EOM}
 \end{equation}
 describing the dynamics of the expectation value $ \langle O\rangle$ of an operator $O$. We model losses and dephasing via the common Lindbladian
\begin{equation}
 \mathcal L_L(O)=\frac{\i}{2}\left(L^\dagger[O,L]+[L^\dagger,O]L\right).
 \label{eq:LBH}
\end{equation}
 We include photon losses from the cavity with the Lindblad operator $L=\sqrt{\hbar\kappa}b$ and polarization decay -- also referred to as pure dephasing -- via $L=\sqrt{\hbar\gamma_{i,j}}\sigma_{i,j}$ for $i>j$ \cite{schneebeli2010zeno}. Here, we have introduced the inversion operator $\sigma_{i,j}=P_{i,i}-P_{j,j}$. Assuming $\gamma_{i,j}=\gamma_{j,i}$, all the dephasing contributions are the same for the four configurations. Radiative decay, i.e., recombination of the excited levels into the ground states, can be included for the configuration-specific optically allowed transitions $P_{i,j}$ with $E_j>E_i$ via $L=\sqrt{\hbar r_{i,j}}P_{i,j}$. Explicitly, the decay term for $O=b^{(\dagger)}$ is given by
\begin{equation}
 \sum_{L}\mathcal L_{L}(b^{(\dagger)})=-\frac{\i\hbar\kappa}{2}b^{(\dagger)}
\end{equation}
and the pure-dephasing term for $O=P_{m,n\neq m}$  by
\begin{align}\label{eq:poldeph}
 \left.\sum_{L}\mathcal L_{L}(P_{m,n})\right|_\mathrm{pure}&=-\frac{\i\hbar}{2}\sum_{i>j}\gamma_{i,j}(2\delta_{i,n}\delta_{j,m}+2\delta_{i,m}\delta_{j,n}\\ \notag
                                                                                                     &\quad+\delta_{i,m}+\delta_{j,m}+\delta_{i,n}+\delta_{j,n})P_{m,n}.
\end{align} 
We note that the inclusion of radiative contributions to the decay is straightforward but complicates a direct comparison of the three-level systems of different type. A detailed discussion is given in Appendix \ref{app:raddec}. 

We use the cluster-expansion method \cite{kira2006many,berger2014quantum} to deal with the many-particle hierarchy problem resulting from Eq.~\eqref{eq:EOM}, i.e., in normal order, we expand all one-, two-, three-operator and higher-order expectation values into their \emph{correlated clusters}. These correlations are named singlets, doublets, triplets, quadruplets and so forth, corresponding to their respective interaction hierarchy, and are defined via a recursive factorization scheme. We indicate correlations of an operator quantity $\langle O\rangle$ with double brackets,  $\llangle O\rrangle$. The singlets are identical with their full expectation value, here $\llangle P_{i,j}\rrangle=\langle P_{i,j}\rangle$ and $\llangle b^{(\dagger)}\rrangle=\langle b^{(\dagger)}\rangle$. For the doublets appearing in our calculations, we find
 \begin{align}
 \langle b^\dagger b\rangle&=\langle b^\dagger\rangle\langle b\rangle+\llangle b^\dagger b\rrangle=\llangle b^\dagger b\rrangle,\notag\\
 \langle P_{i,j} b^\dagger\rangle&=\langle P_{i,j}\rangle\langle b^\dagger\rangle+\llangle P_{i,j} b^\dagger\rrangle=\llangle P_{i,j} b^\dagger\rrangle.
\end{align}
Note that under our excitation conditions (the laser polarization is assumed to be orthogonal to the cavity polarization) the photon singlets $\langle b^{(\dagger)}\rangle$ are always zero.  Analogously, we find for the triplets
\begin{align}
 \langle P_{i,j} b^\dagger b\rangle&=\langle P_{i,j}\rangle\llangle b^\dagger b\rrangle+\llangle P_{i,j} b^\dagger b\rrangle
\end{align}
and for the quadruplets 
\begin{align}
 \langle P_{i,j} b^\dagger bb\rangle&=2\llangle P_{i,j}b\rrangle\llangle b^\dagger b\rrangle+\llangle P_{i,j} b^\dagger bb\rrangle,\notag\\
 \langle P_{i,j} b^\dagger b^\dagger b\rangle&=2\llangle P_{i,j}b^\dagger\rrangle\llangle b^\dagger b\rrangle+\llangle P_{i,j} b^\dagger b^\dagger b\rrangle.
\end{align}
In the following, we implement the singlet-doublet-triplet (SDT) approximation and neglect the quadruplets and all higher-order correlations. 

 To compute the \emph{full} emission dynamics as a reference, we can directly solve the von-Neumann equation for the density operator $\rho$,
 \begin{equation}
  \i\hbar\ddt \rho=[H,\rho]+\sum_L\tilde{\mathcal L}_L(\rho),
\label{eq:vNE}
 \end{equation}
 where $\tilde{\mathcal L}_L(\rho)$ is the Lindbladian Eq.~\eqref{eq:LBH} in the Schr\"odinger picture,
 \begin{equation}
 \tilde{\mathcal L}_L(\rho)=\frac{\i}{2}\left(2L\rho L^\dagger-\rho L^\dagger L-L^\dagger L\rho \right).
 \label{eq:LBS}
\end{equation}
 The expectation value of an operator $O$ is then defined via taking the trace with the density operator, $\langle O \rangle=\tr\left(\rho O\right)$. This exact appropach is used to verify the cluster-expansion approximation described above, see also Appendix \ref{app:verify}. In the following, we always assume that all the three-level systems are initialized in state 1 without any initial electronic polarizations and without any photons inside the cavity.

\section{Results}

\begin{figure}[t!]
 \centering
 \includegraphics[width=0.47\textwidth]{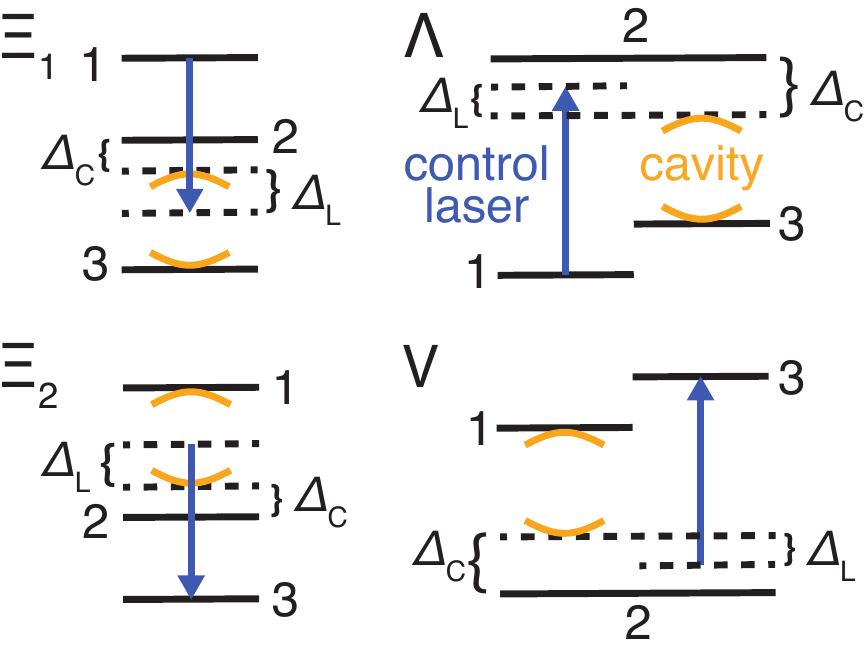}
 \caption{Illustration of cavity,  $\DC$, and control-laser, $\DL$, detunings according to Eqs. \eqref{eq:DL} and \eqref{eq:DC}. The yellow brackets mark the cavity frequency and the blue arrows the laser frequency. The reference points for the respective detunings are highlighted by the dashed lines.}
\label{fig:detunings}
\end{figure}

To thoroughly analyze the Raman-emission dynamics -- which is strongly affected by light-induced energy shifts of the Stark type -- we allow the control laser to be tunable around the bare Raman-resonance condition via an adjustable detuning $\DL$ and define the excitation frequency $\wL$ as
\begin{equation}
\hbar\wL=|\hbar(\omega_1-\omega_3-\wC)|+\DL.
\label{eq:DL}
\end{equation}
We may further assume $|\omega_1-\omega_3|<\wC$ for the V- and $\Lambda$-configurations.
Moreover, we introduce the detuning of the cavity from the two-level subsystem as 
\begin{equation}
\DC=\hbar(\omega_i-\omega_j-\wC)
\label{eq:DC}
\end{equation}
 with $(i,j)=(1,2)$ for the V-type and $(2,3)$ for the $\Lambda$-type system, hence $\omega_i-\omega_j>0$. Both definitions of these detunings are sketched in Fig. \ref{fig:detunings}. In this context, we further assume typical energy scales of semiconductor-QD-based systems where energy differences of optically allowed transitions are on the order of $1-2$ eV.
 
As detailed in the previous section, we derive the full sets of equations of motion for the two system classes within the SDT approximation, the equations are explicitly given in Appendix \ref{app:SDTeqs}. Whenever we use the $\pm$-notation the upper sign refers to the $\Lambda$- or  V-configuration of the respective system class and the lower sign to the  $\Xi$-configurations. To express  the appearing energy differences by the above defined  detunings $\DL$ and $\DC$, we go into the rotating frame of the light field. For a quantity $Q$ the envelope part $Q_\env$ is given via $Q(t)=Q_\env(t)\e^{\pm\i\wL t}$.
 
 To show that $\Lambda$- and V-type systems exhibit different Raman-emission behaviour, in the first instance we analyze and compare their equations of motion of the Raman-photon-assisted polarization 
\begin{equation}
	R\equiv\drei
\end{equation}
 which is a doublet correlation. To discriminate between the photon-assisted polarization $R$ and the photon-assisted polarization of the respective cavity-two-level subsystem, $\varPi\equiv\zweiL$ ($\Lambda$-type) and $\zweiV$ (V-type), we refer to $R$ as \emph{Raman correlation} or  \emph{Raman polarization} and to $\varPi$ commonly as \emph{photon-assisted polarization}. Moreover, for a quantity $Q$, we will indicate the system type $T$ by the respective superscript, $Q^T$ with  $T=\Lambda$ or V. 

\subsection{Raman-correlation equation}
The Raman correlation of $\Lambda$-type systems obeys the equation of motion (see Eq.~\eqref{eq:SDTL3})
\begin{align}\label{eq:RamanL}
 \ddt R_\env^\Lambda&=-\i\left(\pm\frac{\DL}{\hbar}-\i\frac{\kappa+\varGamma_{3,1}}{2}\right)R_\env^\Lambda-\frac{\i\Oenv}{\hbar}\varPi^\Lambda\notag\\
                     &\quad+\frac{\i g}{\hbar}P_\env^\Lambda\left(1+N^\Lambda\right)+\frac{\i g}{\hbar}C_\env^\Lambda
 \end{align}
 and the Raman correlation of V-type systems obeys the equation (see Eq.~\eqref{eq:SDTV3})
 \begin{align}\label{eq:RamanV}
 \ddt R_\env^\V&=-\i\left(\pm\frac{\DL}{\hbar}-\i\frac{\kappa+\varGamma_{3,1}}{2}\right)R_\env^\V+\frac{\i\Oenv}{\hbar}\varPi^\V\notag\\
                &\quad-\frac{\i g}{\hbar}P_\env^\V N^\V-\frac{\i g}{\hbar}C_\env^\V.
 \end{align}
The Raman polarization is a coherent correlation as it only exists in the presence of the control-laser field. Here, we have introduced the dephasing constants $\varGamma_{i,j}$ for the polarizations -- in general containing both radiative decay and pure dephasing -- which are explicitly given by Eq.~\eqref{eq:poldeph} modified by the respective radiative-loss constant  (see Appendix \ref{app:raddec}).  $N^T$ denotes the cavity-photon population $\eins$ of the $T$-type system and $n_i^T$ the electronic occupation of level $i$, $\langle P_{i,i}\rangle$. $P^T$ is the laser-induced electronic polarization with $P^\Lambda\equiv\langle P_{2,1}\rangle$ and $P^\V\equiv\langle P_{3,2}\rangle$.  In total, we find three fundamental source terms.  In the first line of both equations, the laser field is explicitly entering and converts photon-assisted polarization $\varPi$ into Raman correlation $R$. The first term in the second line describes how the laser-induced polarization $P$ acts as a source of Raman correlation. Furthermore, there is a triplet source $C$ which is a correlation between the laser-induced polarization and the emitted photon-population. Equations \eqref{eq:RamanL} and \eqref{eq:RamanV} are formally \emph{exact} within the SDT approximation because Raman correlation as a doublet correlation only couples to triplet correlations  according to the corresponding hierarchy in the Heisenberg equation of motion.
 
Comparing the source terms of Eqs. \eqref{eq:RamanL} and \eqref{eq:RamanV}, we identify structural and, due to the initial condition, further differences. In $V$-type systems, $P^\V$ only enters through the feedback with the cavity field $N^\V$ via $P_\env^\V N^\V$. In $\Lambda$-type systems $P^\Lambda$ does not only enter via backaction of the cavity field but also  appears as a separate source term via $P_\env^\Lambda\left(1+N^\Lambda\right)$. Another difference between the system classes is produced by the different dynamics of the $\varPi$-term. Since V-type systems spontaneously decay via the cavity once the initial state 1 is occupied, $\varPi^\V$ contains a non-vanishing source term even when no control field is present. This spontaneous incoherent source term is immediately available for the conversion into Raman correlation applying an appropriate control pulse. For comparison, in $\Lambda$-type arrangements  $P^\Lambda$ must be created first, before the build-up of Raman correlation can occur. These observations from the equation structure of the Raman correlation will become more clear later during the analysis of measurable quantities shown in Figs. \ref{fig:decomp}--\ref{fig:emissionprob2}.

Equations \eqref{eq:RamanL} and \eqref{eq:RamanV} can be solved analytically in the quasi-stationary limit for weak continuous-wave (cw) excitation (see Appendix \ref{app:Quasistat}). Many properties derived in this regime are -- at least qualitatively -- conserved also in the limit of non-stationary conditions, e.g., pulsed excitations or excitations causing occupation inversion. In particular, we can detect how Stark shifts affect the Raman-excitation conditions. Namely, for the case of optimum Raman-excitation conditions we can identify a common expression for the optimum-laser detuning in the cw limit ($\varOmega_\env(t)=\varOmega_0$) for both system classes,
\begin{equation}
 \DL^\mathrm{cw}=\pm\frac{\DC\varOmega_0^2}{\DC^2+\left(\hbar\frac{\kappa+\varGamma_{3,2[2,1]}}{2}\right)^2},
\label{eq:DLoptcw}
\end{equation}
for $\Lambda$[V]-type systems. Hence, the sign of the detuning  $\DC$ decides if the control laser introduces a blue or red shift of the photon emission according to energy conservation of the two-photon process, in agreement with earlier investigations of a related situation \cite{brewer1975coherent}.  Furthermore, our above made observations concerning the fundamental differences between the source terms also become obvious in the analytic cw solution. Namely, we find that the $\Lambda$-type Raman correlation scales  like 
\begin{equation}
	|R^\Lambda_\mathrm{cw}|^2\propto\frac{g^2\varOmega_0^2}{(\DC\mp\DL)^2+\left(\hbar\frac{\varGamma_{2,1}}{2}\right)^2}
	\label{eq:RLprop}
\end{equation}
and the V-type Raman correlation like
\begin{equation}
	|R^\V_\mathrm{cw}|^2\propto\frac{g^2\varOmega_0^2}{\DC^2+\left(\hbar\frac{\kappa+\varGamma_{2,1}}{2}\right)^2}.
	\label{eq:RVprop}
\end{equation}
We clearly see that Eq. \eqref{eq:RLprop} contains an additional resonance condition for $\DL$ besides Eq. \eqref{eq:DLoptcw} stemming from the specific $P^\Lambda$-coupling in Eq. \eqref{eq:RamanL}. The condition $\DC\mp\DL=0$ means that the control laser becomes resonant with the electronic transition between levels 1 and 2.
In contrast to that, $R^\V$ mainly scales monotonically with $\varOmega_0$ according to Eq. \eqref{eq:RVprop}. This difference will become clearly  visible in the input-output characteristics of the Raman emission shown in Figs. \ref{fig:emissionprob1} and \ref{fig:emissionprob2}.

So far we can summarize that $\Lambda$- and V-type systems show some significant differences mainly due to a qualitatively different dependence on the control-laser induced energy shifts. This has not been discussed to its full extent in previous studies where these shifts were neglected or their influence underestimated \cite{kuhn1999controlled,kuhn2002deterministic,keller2004continuous,santori2009indistinguishability}.    In a next step, we analogously analyze the source terms of the cavity-photon population $N$. This will allow us to find an accurate and unambiguous definition of  \emph{Raman population} $N_\R$ which will be needed for any further discussion of our systems  as a Raman-based single-photon source.

\subsection{Definition of Raman population}
The cluster-expansion method enables us to establish a reasonable decomposition of the total photon population $N$ created inside the cavity into different terms stemming from different microscopic physical processes present in our systems \cite{berger2014quantum}. For a $T$-type system, the total photon population is given via formal integration of Eqs.~\eqref{eq:SDTL1} and \eqref{eq:SDTV1} by:

\begin{equation}\label{eq:Ntot}
  N^T(t)=\frac{2g}{\hbar}\Im\int_0^t\d t^\prime\e^{-\kappa(t-t^\prime)}\varPi^T(t^\prime).
 \end{equation}
Here, the photon-assisted polarization $\varPi^T$ is analogously obtained from integration of Eq.~\eqref{eq:SDTL2} yielding
 \begin{align}\label{eq:PiL}
 \varPi^\Lambda(t)&=\int_0^t\d t^\prime\e^{-\i\left(\frac{\DC}{\hbar}-\i\frac{\kappa+\varGamma_{3,2}}{2}\right)(t-t^\prime)}\left[-\frac{\i\Oenv(t^\prime)}{\hbar}R_\env^\Lambda(t^\prime)\right.\notag\\
                  &\quad+\frac{\i g}{\hbar}n_2^\Lambda(t^\prime)-\frac{\i g}{\hbar}\sigma_{3,2}^\Lambda(t^\prime)N^\Lambda(t^\prime)-\left.\frac{\i g}{\hbar}\varSigma^\Lambda(t^\prime)\right]
\end{align}
and integration of Eq.~\eqref{eq:SDTV2} yielding
\begin{align}\label{eq:PiV}
   \varPi^\V(t)&=\int_0^t\d t^\prime\e^{-\i\left(\frac{\DC}{\hbar}-\i\frac{\kappa+\varGamma_{2,1}}{2}\right)(t-t^\prime)}\left[\frac{\i\Oenv(t^\prime)}{\hbar}R_\env^\V(t^\prime)\right.\notag\\
                  &\quad+\frac{\i g}{\hbar}n_1^\V(t^\prime)-\frac{\i g}{\hbar}\sigma_{2,1}^\V(t^\prime)N^\V(t^\prime)-\left.\frac{\i g}{\hbar}\varSigma^\V(t^\prime)\right].
 \end{align}
In Eqs. \eqref{eq:PiL} and \eqref{eq:PiV}, the source term in the first line describes the Raman process (denoted as R). In the second line, the first term is a purely electronic source (denoted as E) related to electronic-occupation transfer into or from the upper level of the cavity-two-level subsystem. The second term describes the backaction of the electronic inversion $\sigma^T_{i,j}=n^T_i-n^T_j$ with the created cavity population (denoted as B) and the third term,  $\varSigma^\Lambda=\acht$ and $\varSigma^\V=\sieben$, is a triplet contribution (denoted as T) resulting from correlations between photon population and electronic inversion.
Both equations are \emph{exact} because -- as mentioned in the previous section -- the triplet order is the highest appearing order in the equation of motion of a doublet quantity.

According to the above defined source terms of type $S$, we can eventually specify the photon-population decomposition, Eq. \eqref{eq:decomp0}, yielding
\begin{equation}\label{eq:decompo}
 N^T=\sum_{S=\mathrm{R,E,B,T}}N^T_S.
\end{equation}
Notably, the Raman population is explicitly given by
\begin{align}\label{eq:Ramanpop}
 N_\R^{\V[\Lambda]}(t)&=\Im\bigg\{\frac{[-]2\i g}{\hbar^2}\int_0^t\d t^\prime\int_0^{t'}\d t''\e^{-\kappa(t-t^\prime)}\notag\\
                                     &\quad\times\e^{-\i\left(\frac{\DC}{\hbar}-\i\frac{\kappa+\varGamma_{2,1[3,2]}}{2}\right)(t'-t'')}\Oenv(t'')R_\env^{\V[\Lambda]}(t'')\bigg\}
\end{align}
where expressions in square brackets are for $\Lambda$-type systems. We note that this is an unambiguous and precise definition of Raman-photon population. It can be computed directly from the SDT equations or via the full solution of the von-Neumann equation, Eq. \eqref{eq:vNE} (see Appendix \ref{app:verify}). We note that $N_\R$ defined by Eq.~\eqref{eq:Ramanpop} and other contributions $N_S$
 can become negative while $N=\sum_S N_S$ as a physical population always remains positive. We will discuss and interpret this property below in Section \ref{sec:Genemprob} in the context of the corresponding photon-emission probability. By the introduction of a \emph{generalized} emission probability, we will specify the condition for which the Raman process leads to the emission of a pure single Raman photon or just diminishes the photon emission stemming from other contributing processes.
 
\begin{figure}[t!]
 \centering
 \includegraphics[width=0.47\textwidth]{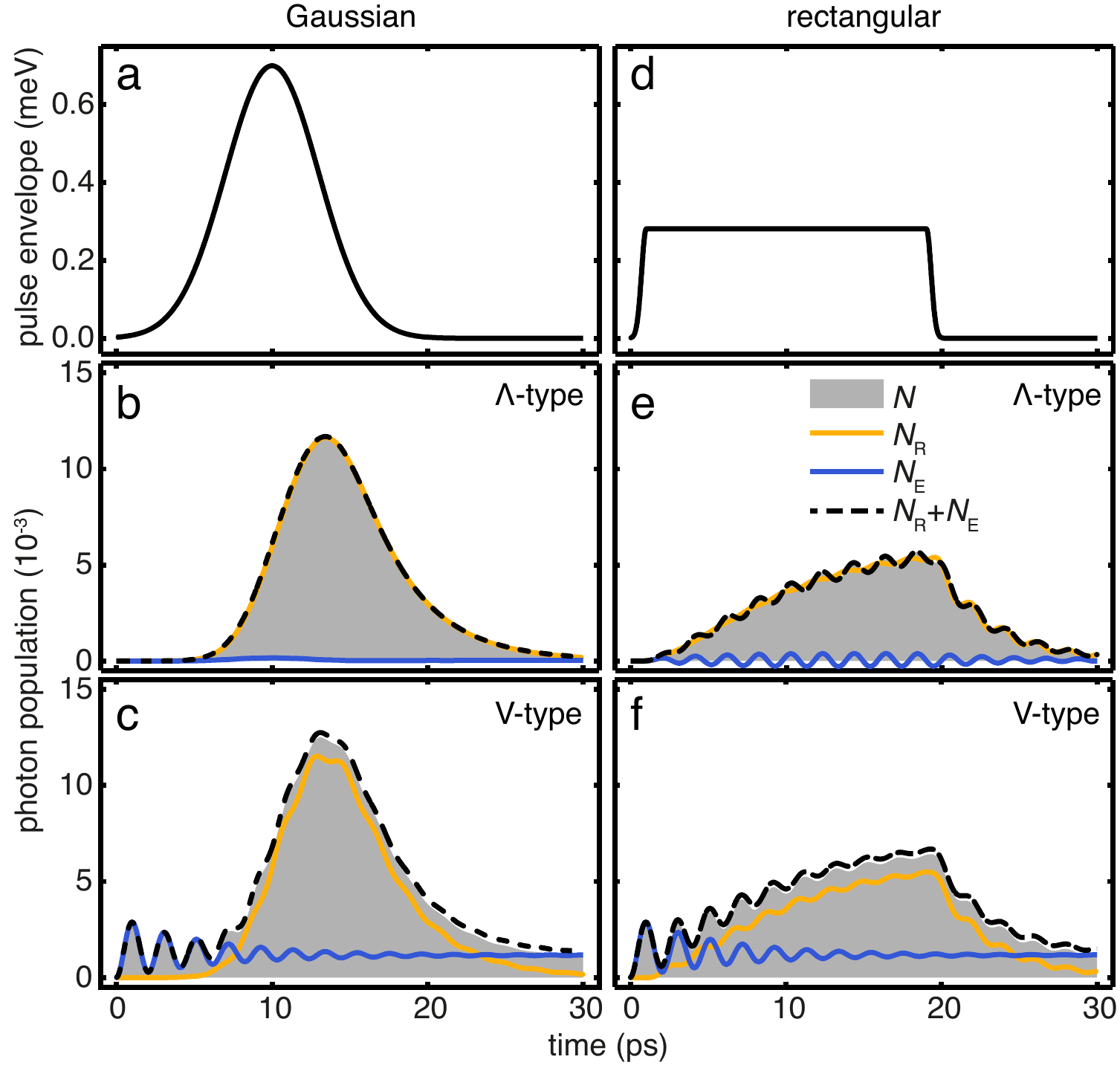}
 \caption{Population decomposition. Shown is (a) the Gaussian and (d) the rectangular  control-pulse profile  $\varOmega_\env$  (black solid line) together with the resulting photon-population dynamics in SDT approximation of (b,e) a $\Lambda$-type system and (c,f) a V-type system  for optimum-Raman conditions. Shown is the full photon population $N$ (shaded area), the Raman population $N_\mathrm R$ (yellow line), the occupation-induced population $N_\mathrm E$ (blue line) and their sum $N_\mathrm R+N_\mathrm E$ (black dashed line). The cavity with $\hbar\kappa=186$ $\upmu$eV is strongly detuned, $\DC=2$ meV, without any spectral overlap with the electronic transition. The other parameters are $g=57$ $\upmu$eV and $\hbar\varGamma_{i,j}=30$ $\upmu$eV. The corresponding optimum-control detuning is $\DL^\opt=\pm0.15$ meV (Gaussian) and $\DL^\opt=\pm0.039$ (rectangular). The choice of the material parameters is compatible with Ref.~\onlinecite{vora2015spin}.}
\label{fig:decomp}
\end{figure}

To give a brief illustration of our decomposition, Eq. \eqref{eq:decompo}, and of the discussions above, we show typical examples of excitation dynamics for both system classes in Fig. \ref{fig:decomp}. We present the SDT photon-emission dynamics for a Gaussian control pulse with a maximum Rabi energy of 0.7 meV [Figs. \ref{fig:decomp}a-\ref{fig:decomp}c]  and the dynamics of a rectangular pulse  of the same pulse area $\theta=5\uppi$ (maximum Rabi energy of 0.281 meV) [Figs. \ref{fig:decomp}d-\ref{fig:decomp}f]. According to our RWA Hamiltonians, the pulse area is given by
\begin{equation}
 \theta=\frac{2}{\hbar}\int_0^\infty\d t\varOmega_\env(t).
\end{equation}
At a first glance, our choice of the two control pulses might seem arbitrary, but they are carefully chosen to highlight that the Raman process exhibits a non-trivial dependence on pulse shape and peak intensity. Both pulse durations are chosen such that the induced Raman-photon populations are completely emitted inside the same time window (see the yellow lines in Fig.~\ref{fig:decomp}). This allows for the most reasonable comparison between both excitation scenarios.  In particular, the rectangular pulse is designed to efficiently pick off the incoherent source (blue lines) of Raman correlation in the V-type case [Figs.~\ref{fig:decomp}c and \ref{fig:decomp}f]. Remarkably, in the context of the Raman-emission probability (see discussion of Fig.~\ref{fig:emissionprob1} below), we will see that, in our example, the rectangular pulse -- which has the same pulse area but a lower maximum-Rabi amplitude -- in both system classes can actually lead to a Raman probability twice as high as for the Gaussian control pulse. We note that in the limit where cavity and laser are spectrally tuned into the vicinity of the electronic resonance  -- which will be necessary for achieving on-demand emission (see discussion of Fig.~\ref{fig:probdldc} later) -- the Raman-emission properties are \emph{not} exclusively determined by the bandwidth of the pulse. Our findings clearly indicate that they show a very complex dependency on the combination of pulse bandwidth, pulse shape, pulse intensity and cavity detuning together with the resulting excitation-induced energy shifts (see Eqs.~\eqref{eq:DLoptcw}-\eqref{eq:RVprop} and Ref.~\onlinecite{breddermann2017shift}). 
In our example, we choose the limit of strong cavity detuning where the emission should be dominated by the Raman process. In $\Lambda$-type systems [Figs. \ref{fig:decomp}b and \ref{fig:decomp}e],  the full emission (shaded area) is completely stemming from the Raman emission (yellow line). Electronic-occupation induced photon population (blue line) hardly contributes. In V-type systems, additionally the initial occupation of level 1 instantaneously generates incoherent photon population (blue line) before the control pulse starts to induce Raman emission [Figs. \ref{fig:decomp}c and \ref{fig:decomp}f]. The sum of both contributions (black dashed line) in all four cases describes the full emission very well. In particular, this example shows that our decomposition definition indeed is physically reasonable.  We note that the rectangular pulse is introduced here to generate a hybrid scenario between a pulsed and a cw scenario. We further note that -- although the emission dynamics are non-stationary -- our quasi-stationary cw approximations will apply and lead to a deeper understanding.  In the following, we will investigate the emission characteristics more systematically in dependence of varying cavity detuning, control detuning and control power. In this regard, we will further demonstrate that the incoherent emission in V-type systems can actually be efficiently converted into Raman population. 

For this purpose, we need a unique measure to characterize and compare the performance of our two system classes, given by the photon-emission probability as discussed in the following section.
 
 \subsection{Generalized emission probability}\label{sec:Genemprob}
 The total photon-emission probability up to time $t$ of our QD-cavity systems can be obtained from
 \begin{equation}\label{eq:Pemit}
  \mathcal{P}(t)=\kappa\int_0^t\d t' N(t')\,,
 \end{equation}
 within the Lindblad formalism. This quantity actually gives the total photon number emitted during the time interval $[0,t]$. Due to our initial condition and assuming that on the  time scale of the photon life time $1/\kappa$ no loop  $1\rightarrow3\rightarrow1$  among the electronic configurations has to be considered (e.g., via nonradiative spin relaxation between the two ground states 1 and 3 in a QD-$\Lambda$-configuration \cite{sweeney2014cavity}), a maximum of one photon can be emitted into the cavity mode during a single excitation cycle. Hence, we identify $\mathcal P=1$ with $100\%$ emission probability such that $0\%\leq\mathcal P\leq 100\%$. The use of our decomposition, Eq.~\eqref{eq:decompo}, in Eq.~\eqref{eq:Pemit} yields
 \begin{equation}\label{eq:Pdec}
  \mathcal{P}(t)=\sum_{S}\kappa\int_0^t\d t' N_S(t')\equiv\sum_{S}\mathcal{P}_S(t)
 \end{equation}
 where in particular the Raman-photon emission probability of type $T$,
 \begin{equation}\label{eq:PRdec}
  \mathcal{P}^T_\R(t)=\kappa\int_0^t\d t' N_\R^T(t'),
 \end{equation}
 is defined via Eq.~\eqref{eq:Ramanpop}. We have to emphasize that $\mathcal P_S$ can become negative or greater than $100\%$ which is, as already mentioned above in the context of the population decomposition, the fingerprint of quantum interference. For that reason, $\mathcal P_S$ has to be understood as a \emph{generalized} emission probability. It can be identical with a \emph{physical} probability in the case where $0\%\leq\mathcal P_S\leq 100\%$ for all sources $S$.
For a chosen cavity detuning and control amplitude, the optimum control detuning $\DL^\opt$ is then defined as the control detuning $\DL$ for which the  Raman-emission probability reaches its maximum,
\begin{equation}
 \DL^\opt=\left.\DL\right|_{\mathrm{max}(\mathcal P_\mathrm R)}.
\label{eq:DLoptnum}
\end{equation}

 We can directly apply our emission-probability decomposition, Eq. \eqref{eq:Pdec}, to our example shown in Fig. \ref{fig:decomp}. For the Gauss-pulse dynamics we find $\mathcal P_\mathrm R^\Lambda(t=30\mbox{ ps})=2.85\%$ as the Raman-emission probability and $\mathcal P_\mathrm E^\Lambda(t=30\mbox{ ps})=0.04\%$ as the emission probability of the electronic contribution [Fig. \ref{fig:decomp}b] and  $\mathcal P_\mathrm R^\mathrm V(t=30\mbox{ ps})=2.82\%$ and $\mathcal P_\mathrm E^\mathrm V(t=30\mbox{ ps})=1.04\%$ [Fig. \ref{fig:decomp}c]. For the rectangular-pulse dynamics we calculate $\mathcal P_\mathrm R^\Lambda(t=30\mbox{ ps})=2.2\%$ and $\mathcal P_\mathrm E^\Lambda(t=30\mbox{ ps})=0.04\%$ [Fig. \ref{fig:decomp}e] and  $\mathcal P_\mathrm R^\mathrm V(t=30\mbox{ ps})=2.25\%$ and $\mathcal P_\mathrm E^\mathrm V(t=30\mbox{ ps})=1.04\%$ [Fig. \ref{fig:decomp}f]. That means, in all cases the emission probability of a Raman photon is only $2-3\%$ and far away from on-demand behaviour. We therefore have to systematically analyze how and under which conditions on-demand behaviour can be achieved or approached.

For a consistent fundamental  comparison of $\Lambda$- and V-type systems based on simulations -- where we want to create \emph{identical} conditions for both system classes --, we neglect radiative decay of the electronic occupations which does not depend on the system class ($\Lambda$-type or V-type) but on the actual  \emph{configuration} ($\Xi$, $\Lambda$ or V) of the respective system. Since radiative decay only modifies the electronic damping constant $\varGamma_{3,1}$ and does \emph{not} change the structure of the Raman equations, Eqs. \eqref{eq:RamanL} and \eqref{eq:RamanV},  this will hardly influence the physical observations from our numerical results presented in the following.
In particular, neglecting radiative decay is valid for pulsed excitations where the duration of the control pulse is significantly shorter than the radiative lifetime of the occupations\cite{heinze2015quantum,breddermann2016tailoring}. A detailed discussion of the influence of radiative decay is given in Appendix~\ref{app:raddec}.
 
  \begin{figure}[t!]
 \centering
 \includegraphics[width=0.47\textwidth]{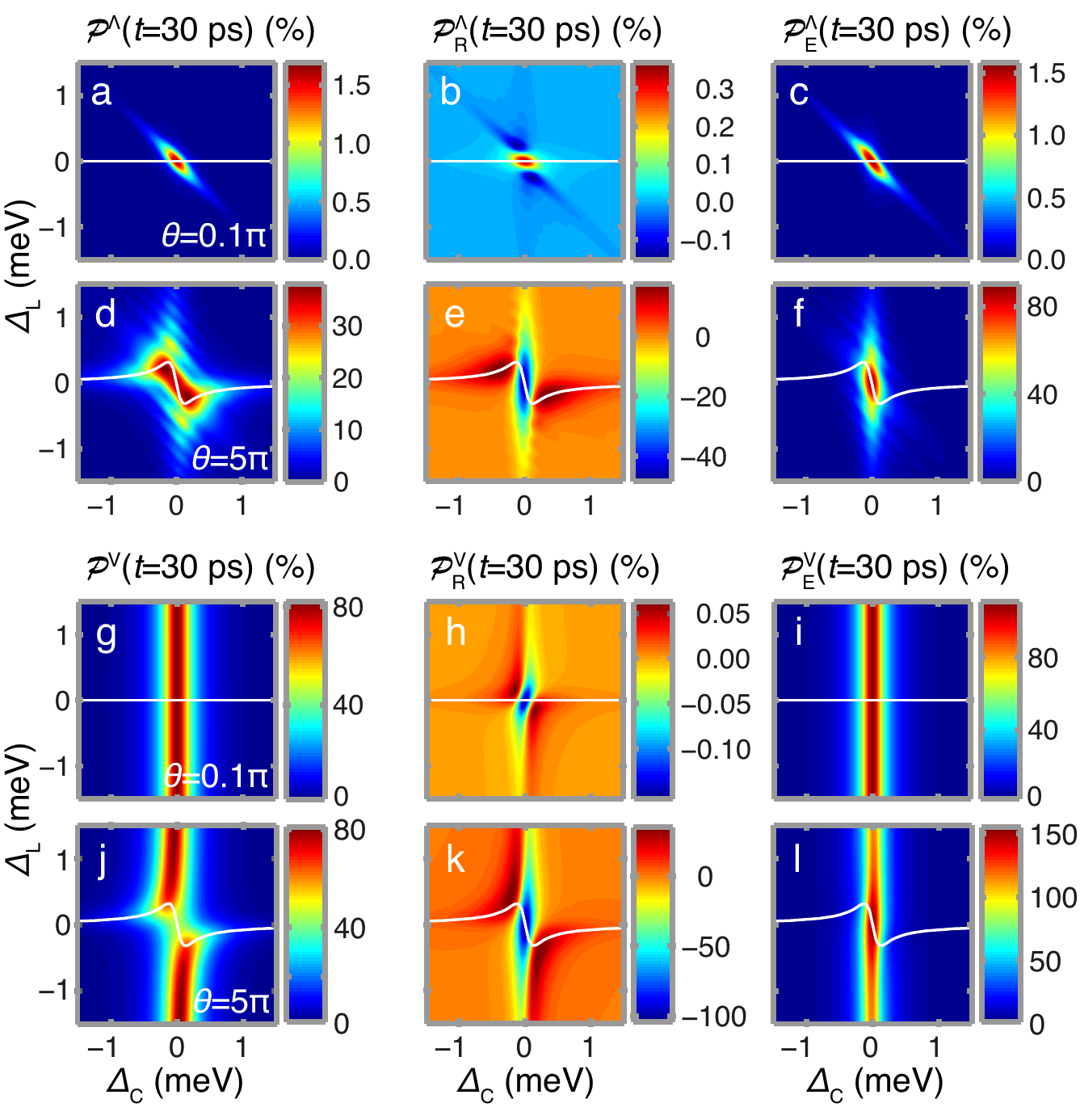}
 \caption{Emission probability for varying control- and cavity detuning. Shown are the emission probabilities of a $\Lambda$-type system (a-f) and  of a V-type system (g-l). The first column shows the full emission probability $\mathcal P^T(t=30\mbox{ ps})$ together with its decomposition into Raman $\mathcal P^T_\mathrm R$ (second column) and electronic part $\mathcal P^T_\mathrm E$ (third column) for the rectangular pulse profile shown in Fig. \ref{fig:decomp}.  The first and third row show a weak-excitation scenario with a $0.1\uppi$-pulse, and  the second and fourth row a strong-excitation scenario with a $5\uppi$-pulse. The analytic optimum-Raman condition obtained from Eq. \eqref{eq:DLoptcw} is shown by the white line, respectively. We explicitly show the results for the $\Xi$-configurations here. The results for $\Lambda$- and V-configurations can be obtained by mirroring the patterns along the $\DL$=0-axis. The system parameters are the same like in Fig. \ref{fig:decomp}.}
\label{fig:probdldc}
\end{figure}

Figure \ref{fig:probdldc} shows the emission probabilities for varying control detuning $\DL$ and cavity detuning $\DC$ of two representative excitation scenarios for a $\Lambda$-type (upper block) in comparison with a V-type system (lower block), one for a weak rectangular pulse ($\theta=0.1\uppi$) and one for a strong rectangular pulse ($\theta=5\uppi$).  
The probabilities are calculated at time $t=30$ ps, about 10 ps after the presence of the control pulse such that the Raman population has completely been emitted according to Fig. \ref{fig:decomp}. We present the full emission probability $\mathcal P$ (first column) and the two most important components of the decomposition, the Raman-emission probability $\mathcal P_\mathrm R$ (second column) and the electronic emission probability $\mathcal P_\mathrm E$ (third column). In all panels we plot the analytic cw result for the light-induced energy shift, Eq. \eqref{eq:DLoptcw}, as a guide to the eyes. The full emission is approximately represented by the sum of the Raman and electronic part. We observe that especially the weak-excitation scenario -- where Stark shifts are hardly visible -- of a $\Lambda$-type system [panels a-c] looks completely different than the one of a V-type system [panels g-i]. The main difference is that we find a small positive Raman probability ($<0.4\%$) in the $\Lambda$-type system [panel b] for the case of a resonant cavity, $\DC\approx0$, in combination with the bare Raman-resonance condition, $\DL\approx0$. In contrast to that, in the same region of the Raman map, the corresponding V-type system shows a purely negative Raman (generalized) emission probability, hence the Raman process does not generate any photon but causes destructive interference diminishing the emission stemming from the electronic source [panel i]. Additionally, we observe a further negative region in the $\Lambda$-type Raman map in Fig. \ref{fig:probdldc}b along the diagonal $\DL=-\DC$ according to our observations from Eq. \eqref{eq:RLprop}. As can be seen in Fig. \ref{fig:detunings}, along this diagonal the control laser is always resonant with the electronic transition between states 1 and 2. Obviously, the V-type system [Fig. \ref{fig:probdldc}h] is not sensitive to this resonance, in agreement with Eq. \eqref{eq:RVprop}. This is a direct consequence of the different electronic-polarization coupling observed in the Raman equations, Eqs. \eqref{eq:RamanL} and \eqref{eq:RamanV}. The orientation of the $\Lambda$-type electronic emission and total emission also follow the diagonal. The orientation of the V-type electronic emission follows the cavity-resonance line $\DC=0$ because the fully occupied initial state can efficiently decay via the resonant cavity. In the case of strong excitation the Raman patterns of both system classes [panels e and k] become more similar, but their maxima are located differently. The analytic dependency shown by the white line especially describes the optimum line of the $\Lambda$-type Raman map very well. Obviously, Raman interference occurs in both classes in the region characterized by the extremal points of the white line, namely for 
\begin{equation}
 |\DC|\leq\hbar\frac{\kappa+\varGamma_{3,2[2,1]}}{2}
\label{eq:interfcond}
\end{equation}
in a $\Lambda$[V]-type system. Hence, the critical detunings for interference are only determined by the cavity losses, $\kappa$, and the total electronic-dephasing rate, $\Gamma_{i,j}$, of the two-level-cavity subsystem. The corresponding control detunings of the extremal point are dependent on the pulse amplitude, 
\begin{equation}
|\DL|=\frac{\varOmega_0^2}{\hbar(\kappa+\varGamma_{3,2[2,1]})}.
\end{equation}
At this point we want to emphasize that the maxima observed in the full emission [panels d and j] are \emph{not} identical with the maxima of the Raman contribution. Notably, in the V-type system, even when Stark shifts are hardly visible [panel h], the light-induced effects are nevertheless contributing via the interference condition, Eq. \eqref{eq:interfcond}. Interestingly, V-type Raman emission is quite insensitive against the choice of the control detuning over a  range of almost 1 meV for the case of a near-resonant cavity [panels h and k]. Here the Raman emission exhibits a pronounced shoulder due to electronic-to-Raman emission conversion. Corresponding to this efficient conversion, the maximum stripe  of the electronic contribution observed in panel l is much narrower in comparison with the stripe of the weak-excitation scenario [panel i]. This conversion effect is absent in $\Lambda$-type systems.

We do not show the emission maps for the corresponding Gauss-pulse scenario or for enhanced coupling strength $g$, because they qualitatively look very similar. We will discuss these scenarios again below in the context of input-output characteristics.

According to our observations from Figs. \ref{fig:probdldc}d-\ref{fig:probdldc}f, we can now state that the interference dip observed in the spectrum, shown in Fig. \ref{fig:interfer}, actually stems from the Raman process diminishing the emission from the electronic source. Hence, we observe destructive interference between the direct two-photon Raman path $1\stackrel{\varOmega^{(\star)} b^\dagger}{\longrightarrow}3$ -- exclusively mediated by the Raman correlation -- and the path explicitly involving the intermediate state $1\stackrel{\varOmega^{(\star)}}{\longrightarrow}2\stackrel{b^\dagger}{\longrightarrow}3$ via electronic-occupation transfer. Note that the interference feature can be masked in the spectrum if the dip is compensated by the emission from other channels, as can also be seen in Fig. \ref{fig:interfer} for longer measurement durations. This can especially happen in V-type systems where the resonant emission is strongly dominated by the electronic source [Figs. \ref{fig:probdldc}g-\ref{fig:probdldc}l].
Only a detailed decomposition is able to discriminate the different contributing emission channels.

\subsection{Input-output characteristics}

\begin{figure}[t!]
 \centering
 \includegraphics[width=0.47\textwidth]{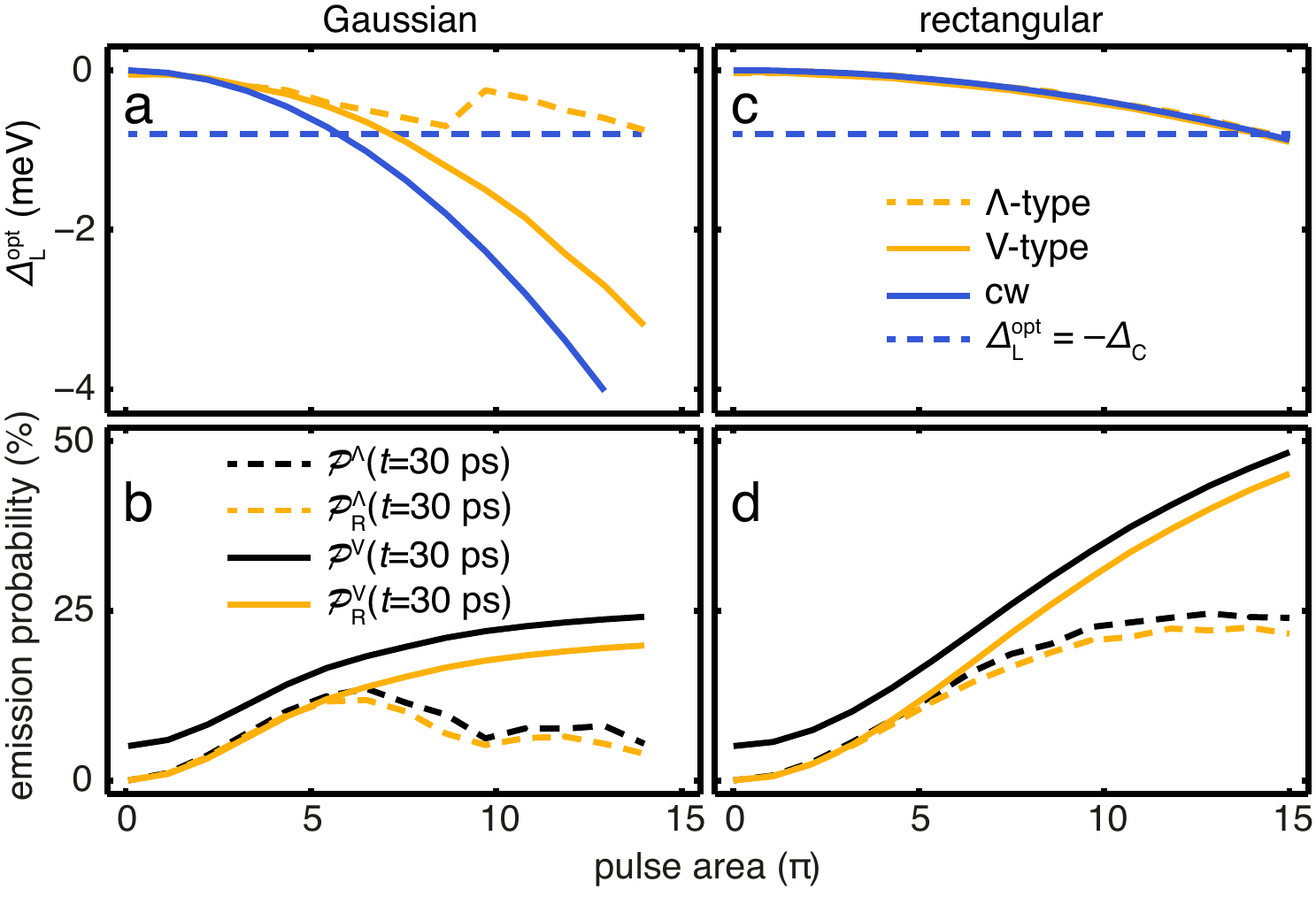}
 \caption{Optimum-emission probability for varying pulse area in the off-resonant-cavity regime. Shown is the optimum-control detuning together with the corresponding emission probability for varying pulse area  $\theta$ of both system classes for the Gaussian pulse (a,b) and the rectangular pulse (c,d). Panels a and c  show the numerically calculated optimum detuning  of the $\Lambda$-type (yellow dashed lines) vs.~the V-type system (yellow solid lines). For comparison, we plot the analytic result (blue solid line), Eq. \eqref{eq:DLoptcw}, and the $\Lambda$-type resonance condition $\DL^\opt=-\DC$ (blue dashed lines), according to Eq, \eqref{eq:RLprop}. In panels b and d, we show the total, $\mathcal{P}(t=30\mbox{ ps})$, (black lines) and the Raman-emission probability, $\mathcal{P}_\R(t=30\mbox{ ps})$, (yellow lines) for the $\Lambda$-type (dashed lines) and the V-type system (solid lines). The parameters are $\DC=0.8$ meV, and as before, $\hbar\kappa=186$ $\upmu$eV, $g=57$ $\upmu$eV, $\hbar\varGamma_{i,j}=30$ $\upmu$eV. We again explicitly show the results for the $\Xi$-configurations. For the $\Lambda$- and V-configurations, panels (b) and (d) remain unchanged, we only get a sign flip of the control detunings in panels (a) and (c). }
\label{fig:emissionprob1}
\end{figure}

Although the Raman maps can have their absolute maxima in the limit of a near-resonant cavity, the efficiency of the Raman process compared with the total output is higher for cavities with intermediate detunings where $\mathcal P\approx\mathcal P_\mathrm R$. For that reason we will now turn our attention to the input-output characteristics for the case where cavity and electronic resonance overlap only moderately. For our chosen system parameters, this is the case for about $|\DC|=0.8$ meV. We vary the control amplitude, i.e., vary the pulse area, and we calculate the corresponding maximum-Raman emission via the optimum-control detuning, Eq. \eqref{eq:DLoptnum}.  We present the input-output characteristics for pulse areas up to $\theta=15\pi$ to highlight the physical differences between the two system classes in a better way. We note that  the range up to $\theta\approx10\pi$ should be experimentally relevant\cite{stufler2006two}.

Figure \ref{fig:emissionprob1} shows a typical example of the optimum-Raman characteristics induced by our Gauss [Figs. \ref{fig:emissionprob1}a and b] and our rectangular pulse [Figs. \ref{fig:emissionprob1}c and d]. The upper panels show the numerically and analytically calculated optimum-control detunings and the lower panels the corresponding total and Raman-emission probabilities, again taken 10 ps after the control pulse. We use our standard system parameters which represent an intermediate cavity quality as experimentally used in Ref.~\onlinecite{vora2015spin}. Both scenarios have in common that the $\Lambda$-type system is a very pure Raman-photon source where the Raman contribution can be near identical with the total emission probability. As already discussed above, the V-type system exhibits a small background stemming from the incoherent electronic source which is also  present without control pulse. This can be seen as a small offset at $\theta=0$ (black solid lines in Figs. \ref{fig:emissionprob1}c and d). The big difference of the input-output characteristic lines is that the V-type system exhibits a quite monotonically increasing behaviour even in the regime of very strong pulses whereas for the $\Lambda$-type system we observe a saturation behaviour for the rectangular case or even a break-down for the Gauss pulse at $\theta\approx6\uppi$. We further observe that the rectangular pulse -- representing a longer coupling duration in the chosen time window -- strongly enhances the output compared with the Gauss-pulse scenario\cite{breddermann2017shift}, here by a factor of two (although the peak Rabi energy of the rectangular pulse is lower while both pulse areas are equal). It further induces a more efficient conversion of spontaneous background emission into Raman emission which can be observed if we compare the smaller difference between the solid lines in Fig. \ref{fig:emissionprob1}d with the difference observed in Fig. \ref{fig:emissionprob1}b. The optimum-control detunings of the rectangular pulse are nearly perfectly described by the cw approximation [panel b] and they are much smaller  than the detunings  for the case of the Gauss pulse [panel a]. Nevertheless, for the Gauss pulse, the sign and the order of magnitude  of the V-type optimum shift are still well described by our cw expression. {But we can conclude that the optimum shift is not only determined by the peak-Rabi energy $\varOmega_0$ of the Gaussian pulse -- obviously, the Gaussian \emph{shape} reduces $|\DL^\opt|$ in comparison with the corresponding cw result [panel a]. Hence, the optimum-laser detuning, $\DL^\opt$, to achieve optimum-Raman emission actually is explicitly pulse-shape dependent\cite{breddermann2017shift}.} The strong deviation of the numerically calculated  $\Lambda$-type Gauss-pulse induced shift (yellow dashed line in panel a) from the parabolic shape is a consequence of the $\Lambda$-type resonance condition Eq. \eqref{eq:RLprop} shown by the blue dashed line. This leads to the observed discontinuous behavior of the optimum shift $\DL^\opt$ and also of the corresponding Raman characteristic line: once the laser comes close to resonance with levels 1 and 2 via $\DC\mp\DL=0$, the Raman-photon emission breaks down, even beyond the resonance line. Our optimization condition Eq. \eqref{eq:DLoptnum} then detects neighbored maxima of the Raman patterns which start to split with increasing pulse area. Note that the shifts induced by the rectangular pulse do \emph{not} cross the resonance line within the chosen intensity range leading to a higher quantum yield. Hence, we can state that V-type systems show a more robust performance in contrast to $\Lambda$-type systems for increasing excitation intensities.

\begin{figure}[t!]
 \centering
 \includegraphics[width=0.47\textwidth]{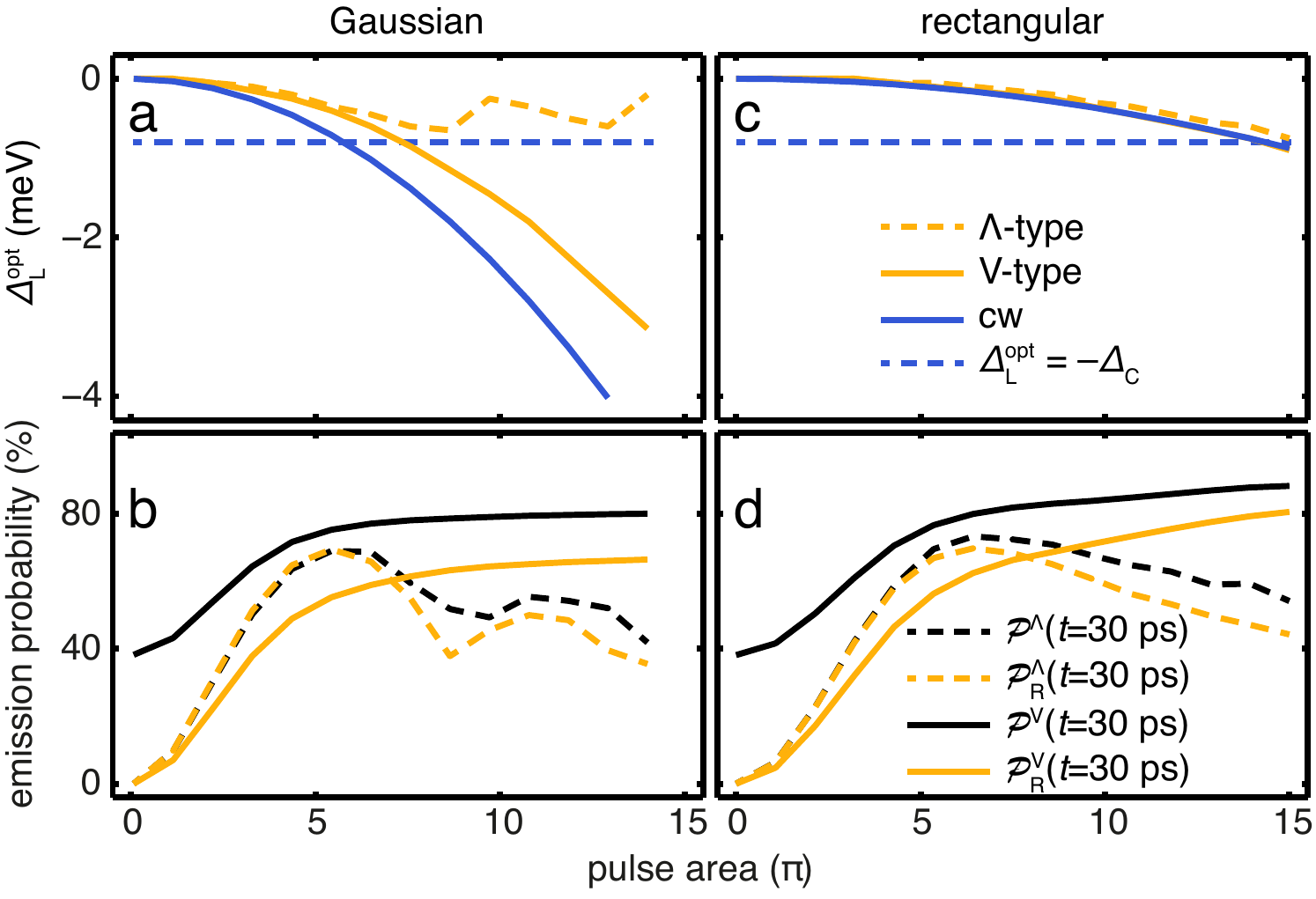}
 \caption{Optimum-emission probability for enhanced light-matter coupling. We present the results for the same situation  as in Fig. \ref{fig:emissionprob1} but with enhanced coupling strength between cavity photons and QD, $g=\hbar\kappa=186$ $\upmu$eV.}
\label{fig:emissionprob2}
\end{figure}

Since we expect a higher quantum yield for enhanced photon-QD coupling $g$ according to Eqs. \eqref{eq:RLprop} and \eqref{eq:RVprop}, we repeat our calculations presented in Fig. \ref{fig:emissionprob1} for enhanced coupling strength, $g=\hbar\kappa=186$ $\upmu$eV. The result is shown in Fig. \ref{fig:emissionprob2} and we observe that the output for both system classes actually reaches roughly $80\%$, even for the Gaussian control pulse. For both control pulses, the $\Lambda$-type system is very efficient around $\theta=6\uppi$, up to this pulse area it even shows a higher Raman yield than the V-type system. While the V-type system operates quite inefficiently below $\theta\approx5\pi$ since the incoherent background is dominating (roughly $40\%$), it shows an increasing or slowly saturating Raman-characteristic line for increasing pulse areas. In contrast to that the $\Lambda$-type input-output line always reveals intensity-induced damping. Figure \ref{fig:emissionprob2}d again demonstrates that incoherent photon emission can be transformed into Raman-photon population: while at $\theta=0$ we find $\mathcal P^\mathrm V-\mathcal P_\R^\mathrm V=40\%$ for the incoherent off-set, we end up with $\mathcal P^\mathrm V-\mathcal P_\R^\mathrm V=5\%$ for $\theta=15\uppi$.

\begin{figure}[t!]
 \centering
 \includegraphics[width=0.47\textwidth]{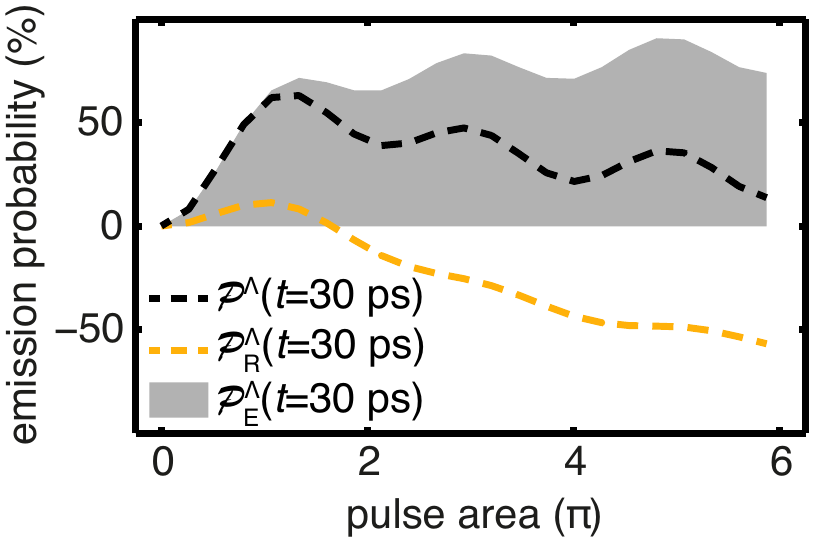}
 \caption{Rabi oscillations in a $\Lambda$-type system. Shown is the total emission probability $\mathcal P^\Lambda$ (black dashed line) together with the Raman probability $\mathcal P_\mathrm R^\Lambda$ (yellow dashed line) and the electronic probability $\mathcal P_\mathrm E^\Lambda$ for the resonant Raman process, $\DC=\DL=0$. The material and excitation parameters correspond to the scenario of Fig. \ref{fig:probdldc}. }
\label{fig:rabi}
\end{figure}
Finally, we come back to the resonant Raman process where both control laser and cavity are resonant with the respective electronic transition, $\DC=\DL=0$. Here, we focus only on $\Lambda$-type systems because -- as discussed before -- V-type systems do not emit Raman photons in the sense of our definition in the resonant limit. Figure \ref{fig:rabi} shows the input-output characteristics of a $\Lambda$-type system for varying pulse area. We  observe damped Rabi oscillations of the total emission $\mathcal P^\Lambda$ (black dashed line).  Obviously, these oscillations stem from the electronic source $\mathcal P_\mathrm E^\Lambda$ (shaded area) related to the path $1\stackrel{\varOmega^{(\star)}}{\longrightarrow}2\stackrel{b^\dagger}{\longrightarrow}3$ which is the dominant positive contribution. The Raman signal $\mathcal P_\mathrm R^\Lambda$ (yellow dashed line) exhibits only a single Rabi oscillation for the $\uppi$-pulse. It becomes negative beyond $\theta=2\uppi$ and decreases quite monotonically for increasing pulse areas. Notably, the Raman correlation is responsible for the \emph{excitation-induced damping}. We find that  this is a \emph{general} phenomenon of the resonant Raman process: efficient occupation transfer from the initial to the intermediate state and the simultaneous build-up of Raman correlation start to counteract for enhanced excitation intensities, because, in $\Lambda$-type systems, both processes share the laser-induced electronic polarization as the only initial source (see also Eqs.~\eqref{eq:SDTL3} and \eqref{eq:P22L}). Interestingly, the $\uppi$-pulse can nevertheless invert the system from the initial to the final state nearly deterministically -- in our example, about $70\%$ of the occupation is transferred (not shown) -- which is commonly introduced as a \emph{spin-flip Raman transition} in $\Lambda$-configurations\cite{he2013indistinguishable}. Obviously, this spin-flip Raman transition is physically different from the Raman-correlation-mediated process. Hence, we conclude that the measured damped Rabi oscillations in a $\Lambda$-system presented in the Supplemental Material of Ref.~\onlinecite{he2013indistinguishable} -- which are induced by \emph{resonant} excitation -- do not originate from the excitation channel $1\stackrel{\varOmega b^\dagger}{\longrightarrow}3$  mediated  by the Raman correlation but purely from the inversion-induced emission. In particular, according to Fig. \ref{fig:rabi}, the damping -- which in Ref.~\onlinecite{he2013indistinguishable} is attributed to acoustic phonons and further non-Markovian dephasing reservoirs -- can already be induced by the Raman correlation itself via destructive interference, even in the absence of phonons. Our findings   also offer an alternative interpretation of the observed Raman-linewidth broadening for vanishing laser detuning: under certain conditions, going into the resonance limit, we do not observe the linewidth-broadening of the Raman process (see Appendix \ref{app:Quasistat}) anymore, but  the  linewidth of the electronic emission which commonly is broader than the Raman linewidth in the cw limit. With the same argument, we may also explain the measured unexpected Raman peak-intensity and broadening behaviour of the resonant Raman process which is discussed and left as an open question in the Supplementary Information of Ref.~\onlinecite{vora2015spin}.

\section{Conclusions}
Using the microscopic cluster-expansion method, we discuss new aspects of cavity-enhanced single-photon generation from optical two-photon Raman processes. We find that either coupling the cavity with the initial or with the final state defines two different classes of three-level-cavity systems which can be characterized by their strongly different Raman-emission dynamics and emission probability. Physically, this difference mainly originates from the different coupling of the control laser with the two-photon Raman process: in contrast to $\Lambda$-type systems, V-type Raman correlation only weakly depends on the control-laser induced electronic polarization. Furthermore, V-type systems exhibit efficient conversion of incoherent emission into coherent Raman emission. We further show that under resonant excitation conditions we may enter the limit of quantum interference where the Raman process does not generate a single Raman photon but acts as a source of destructive interference with other excitation paths.  Hence, the so-called \emph{spin-flip Raman process} in $\Lambda$-configurations -- which allows for deterministic inversion from the initial to final state and which is driven under resonant excitation conditions -- is physically \emph{different} from the non-resonant Raman process -- which additionally allows for all-optical control of the spectral properties of the emitted Raman photon via the control laser.

Based on our model study for a quantum-dot cavity system we predict that the non-resonant Raman protocol -- mediated by the \emph{Raman correlation} introduced in this article -- is indeed able to deliver on-demand photons. This requires appropriately designed control pulses beyond the $\uppi$-pulse regime, where Stark shifts and interference effects have to be taken into account, together with suitably detuned high-quality resonators. The general approach and the fundamental properties reported here can also be transferred to atomic three-level-cavity systems. With the additional microscopic insight obtained in the present work, we are furthermore able to interpret and explain in more detail experimental data from the recent literature\cite{he2013indistinguishable,vora2015spin}. The present work also lays the groundwork for the preparation of related experiments and for the optimization of Raman-process based devices for quantum-optical networking.

\begin{acknowledgments}
 We gratefully acknowledge financial support from the DFG through the research centre TRR142 (project A03) and doctoral training center GRK1464, from the BMBF through Q.com 16KIS0114, and a grant for computing time at $\mbox{PC}^2$ Paderborn Center for Parallel Computing. Stefan Schumacher further acknowledges support through the Heisenberg program of the DFG.
 \end{acknowledgments}

\appendix
\section{Singlet-doublet-triplet equations}\label{app:SDTeqs}
We present the SDT equations, computed from Eq. \eqref{eq:EOM} for a light field of the form $\varOmega(t)=\Oenv(t)\e^{-\i\wL t}$ in the rotating frame. For any quantity $Q$, we define the envelope part via $Q=Q_\env\e^{\pm\i\wL t}$ with the lower sign  for  $\Xi$-configurations and the upper sign for the respectively related $\Lambda$- or V-configuration. Furthermore, we introduce new dephasing constants $\varGamma_{i,j}$ for the polarizations which are explicitly given by Eq. \eqref{eq:poldeph}. For a consistent and reasonable comparison between the system classes by numerical simulations, we do not consider radiative decay. The inclusion of radiative decay and its influence on the Raman process is discussed in the next section.

 \subsection{$\Lambda$-type systems}
The equations of motion for $\Lambda$-type systems without radiative decay are
{\allowdisplaybreaks
\begin{align}
 \ddt\eins&=-\kappa\eins+\frac{2g}{\hbar}\Im\zweiL,\label{eq:SDTL1}\\
 \ddt\zweiL&=-\i\left(\frac{\DC}{\hbar}-\i\frac{\kappa+\varGamma_{3,2}}{2}\right)\zweiL\notag\\
                  &\quad-\frac{\i\Oenv}{\hbar}\dreienv+\frac{\i g}{\hbar}\langle P_{2,2}\rangle\notag\\
                 &\quad-\frac{\i g}{\hbar}\langle \sigma_{3,2}\rangle\eins-\frac{\i g}{\hbar}\acht,\label{eq:SDTL2}\\
\ddt\dreienv&=-\i\left(\pm\frac{\DL}{\hbar}-\i\frac{\kappa+\varGamma_{3,1}}{2}\right)\dreienv\notag\\
                     &\quad-\frac{\i\Oenv}{\hbar}\zweiL+\frac{\i g}{\hbar}\vierLenv\left(1+\eins\right)\notag\\
                &\quad+\frac{\i g}{\hbar}\neunLenv,\label{eq:SDTL3}\\
\ddt\vierLenv&=-\i\left(\frac{-\DC\pm\DL}{\hbar}-\i\frac{\varGamma_{2,1}}{2}\right)\vierLenv\notag\\
                     &\quad-\frac{\i\Oenv}{\hbar}\langle \sigma_{2,1}\rangle+\frac{\i g}{\hbar}\dreienv,\\
\ddt\fuenf&=-\frac{2}{\hbar}\Im\big(\Oenv\vierLenv\big),\\
 \langle P_{2,2}\rangle&=1-\fuenf-\sechs,\label{eq:P22L}\\
\ddt\sechs&=\frac{2g}{\hbar}\Im\zweiL,\\
\ddt\sieben&=-\kappa\sieben+\frac{4}{\hbar}\Im\big(\Oenv\neunLenv\big)\notag\\
                 &\quad-\frac{2g}{\hbar}\left(\eins+\langle \sigma_{2,1}\rangle\right)\Im\zweiL,\\
\ddt\acht&=-\kappa\acht-\frac{2}{\hbar}\Im\big(\Oenv\neunLenv\big)\notag\\
                 &\quad+\frac{2g}{\hbar}\left(2\eins-\langle \sigma_{3,2}\rangle+1\right)\Im\zweiL,\\
\ddt\neunLenv&=-\i\left(\frac{-\DC\pm\DL}{\hbar}-\i\frac{2\kappa+\varGamma_{2,1}}{2}\right)\neunLenv\notag\\
           &\quad+\frac{\i g}{\hbar}\dreienv\eins\notag\\
	        &\quad-\frac{2g}{\hbar}\vierLenv\Im\zweiL\notag\\
	         &\quad-\frac{\i\Oenv}{\hbar}\sieben.
\end{align}
}
Equation \eqref{eq:SDTL3} describes the dynamics of the Raman correlation.

\subsection{V-type systems}
The equations of motion for  V-type systems without radiative decay are
{\allowdisplaybreaks
\begin{align}
 \ddt\eins&=-\kappa\eins+\frac{2g}{\hbar}\Im\zweiV,\label{eq:SDTV1}\\
 \ddt\zweiV&=-\i\left(\frac{\DC}{\hbar}-\i\frac{\kappa+\varGamma_{2,1}}{2}\right)\zweiV\notag\\
                          &\quad+\frac{\i\Oenv}{\hbar}\dreienv+\frac{\i g}{\hbar}\fuenf\notag\\
                 &\quad-\frac{\i g}{\hbar}\langle \sigma_{2,1}\rangle\eins-\frac{\i g}{\hbar}\sieben,\label{eq:SDTV2}\\
\ddt\dreienv&=-\i\left(\pm\frac{\DL}{\hbar}-\i\frac{\kappa+\varGamma_{3,1}}{2}\right)\dreienv\notag\\
                &\quad+\frac{\i\Oenv}{\hbar}\zweiV-\frac{\i g}{\hbar}\vierVenv\eins\notag\\
                &\quad-\frac{\i g}{\hbar}\neunVenv,\label{eq:SDTV3}\\
\ddt\vierVenv&=-\i\left(\frac{-\DC\pm\DL}{\hbar}-\i\frac{\varGamma_{3,2}}{2}\right)\vierVenv\notag\\
                       &\quad-\frac{\i\Oenv}{\hbar}\langle \sigma_{3,2}\rangle-\frac{\i g}{\hbar}\dreienv,\\
\ddt\fuenf&=-\frac{2g}{\hbar}\Im\zweiV,\\
 \langle P_{2,2}\rangle&=1-\fuenf-\sechs,\\
\ddt\sechs&=\frac{2}{\hbar}\Im\big(\Oenv\vierVenv\big),\\
\ddt\sieben&=-\kappa\sieben-\frac{2}{\hbar}\Im\big(\Oenv\neunVenv\big)\notag\\
                 &\quad+\frac{2g}{\hbar}\left(2\eins-\langle \sigma_{2,1}\rangle+1\right)\Im\zweiV,\\
\ddt\acht&=-\kappa\acht+\frac{4}{\hbar}\Im\big(\Oenv\neunVenv\big)\notag\\
                 &\quad-\frac{2g}{\hbar}\left(\eins+\langle \sigma_{3,2}\rangle\right)\Im\zweiV,\\
\ddt\neunVenv&=-\i\left(\frac{-\DC\pm\DL}{\hbar}-\i\frac{2\kappa+\varGamma_{3,2}}{2}\right)\neunVenv\notag\\
                        &\quad-\frac{\i g}{\hbar}\dreienv\left(1+\eins\right)\notag\\
	        &\quad-\frac{2g}{\hbar}\vierVenv\Im\zweiV\notag\\
	         &\quad-\frac{\i\Oenv}{\hbar}\acht.
\end{align}
}
Equation \eqref{eq:SDTV3} describes the dynamics of the Raman correlation.

\section{Discussion of the SDT equations}
In this section we will discuss the SDT equations presented in the previous section, also in comparison with the full solution of the von-Neumann equation Eq. \eqref{eq:vNE}. We will further analyze how the equations change if we include radiative decay of the electronic occupations. 

\subsection{Full vs. cluster-expansion solution}\label{app:verify}
\begin{figure}[t!]
 \centering
 \includegraphics[width=0.47\textwidth]{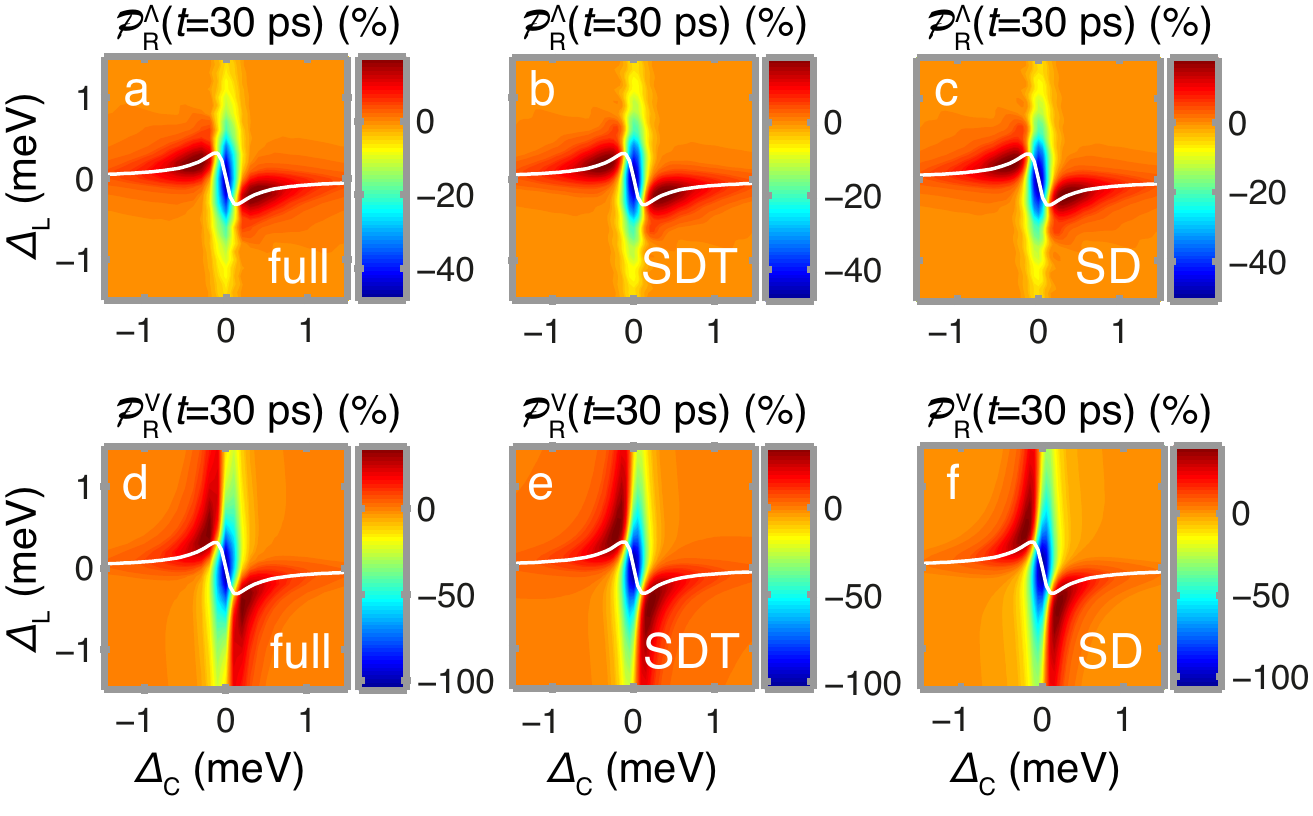}
 \caption{Full vs. singlet-doublet-triplet and singlet-doublet solution. Shown is the Raman-emission probability for the $\Lambda$-type system (a-c) and for the V-type system (d-f). According to Fig. \ref{fig:probdldc}, we compare (first column) the full solution of Eq. \eqref{eq:vNE} for the $5\uppi$-rectangular pulse with (second column) the SDT solution and (third column) with the singlet-doublet (SD) solution. In the SD case, we neglect the triplet correlations and \emph{additionally} switch off the backaction of the cavity field.   We again plot the analytically computed dependency Eq. \eqref{eq:DLoptcw} by the white line.}
\label{fig:Approxcomp}
\end{figure}

We compute the Raman maps using the full solution of the von-Neumann equation as input in our Raman-probability definition $\mathcal P_\mathrm R$, Eq. \eqref{eq:PRdec}. The results are exemplarily shown for the $\Lambda$-type system in Fig. \ref{fig:Approxcomp}a and for the V-type system  in Fig. \ref{fig:Approxcomp}d for the $5\uppi$-pulse scenario from Fig. \ref{fig:probdldc}.
For comparison, we plot the SDT solution [Figs. \ref{fig:Approxcomp}b and \ref{fig:Approxcomp}e]. To further understand the influence of the different source types on the Raman process, we additionally present the singlet-doublet (SD) result [Figs. \ref{fig:Approxcomp}c and \ref{fig:Approxcomp}f] where we neglect the triplet correlations and, in addition to it, the back coupling of the matter with the cavity-light field. 
 On a first view, we can state that the shape of the Raman pattern remains nearly uninfluenced under the SDT and the SD approximation. Since the Raman equation is of the triplet order, in particular the SDT result perfectly agrees with the full solution. In the SD approximation we only detect some minor differences in the quantum yield.
Explicitly, the SD Raman equations without cavity backaction are 
\begin{align}
\ddt\dreienv&=-\i\left(\pm\frac{\DL}{\hbar}-\i\frac{\kappa+\varGamma_{3,1}}{2}\right)\dreienv\notag\\
                     &\quad-\frac{\i\Oenv}{\hbar}\zweiL+\frac{\i g}{\hbar}\vierLenv
\end{align}
for $\Lambda$-type systems and 
\begin{align}
\ddt\dreienv&=-\i\left(\pm\frac{\DL}{\hbar}-\i\frac{\kappa+\varGamma_{3,1}}{2}\right)\dreienv\notag\\
                &\quad+\frac{\i\Oenv}{\hbar}\zweiV
\end{align}
for V-type systems. Hence, we can conclude that -- in contrast to the $\Lambda$-type case -- the V-type Raman process is quite insensitive against the control-laser induced electronic polarization between levels 2 and 3. Its fundamental source is the term describing the conversion of photon-assisted polarization of the two-level cavity subsystem into Raman correlation. This explains the pronounced shoulder around the near-resonant cavity close to the region of destructive interference [Figs. \ref{fig:Approxcomp}d and \ref{fig:Approxcomp}e]. Note that the analytic line misses this shoulder because, in the quasi-stationary regime, we assume a fixed initial occupation which cannot efficiently decay via the near-resonant cavity to be transformed into Raman correlation (see section \ref{app:Quasistat}).  The shoulder feature is not visible in the $\Lambda$-type pattern [Figs. \ref{fig:Approxcomp}a and \ref{fig:Approxcomp}c], because the conversion of electronic emission into Raman correlation is not available in that region of the Raman map.

Nevertheless, if we want to describe the emission dynamics of all quantities accurately, we have to include the full set of SDT equations or use the von-Neumann result.

\subsection{Radiative decay}\label{app:raddec}
Since the inclusion of radiative decay renders a reasonable and consistent comparison of the two different Raman-coupling mechanisms out, we discuss its influence in this subsection separately. Notably, in material systems with longlived occupations or in the case of short excitation durations, radiative decay can be neglected. Radiative decay can explicitly be included in the $\Xi$-configurations via the Lindblad operators $L=\sqrt{\hbar r_{2,1}}P_{2,1}, \sqrt{\hbar r_{3,2}}P_{3,2}$, in the  $\Lambda$-configuration via $L=\sqrt{\hbar r_{1,2}}P_{1,2},\sqrt{\hbar r_{3,2}}P_{3,2}$ and in the V-configuration via $L=\sqrt{\hbar r_{2,1}}P_{2,1},\sqrt{\hbar r_{2,3}}P_{2,3}$. In particular this does not change the structure of the Raman equations, Eqs. \eqref{eq:SDTL3} and \eqref{eq:SDTV3}, only our introduced total dephasing constant $\varGamma_{3,1}$ will be modified in the $\Xi$-systems by $r_{2,1}$ and in the V-system by $r_{2,1}+r_{2,3}$. The Raman process in $\Lambda$-configurations is unaffected by radiative decay, here, in typical realizations with metastable ground states, the spin-dephasing time between the ground states can generate an additional contribution\cite{vora2015spin}. The radiative-decay contributions  for $\varGamma_{2,1}$ are  $r_{3,2}+r_{2,1}$ ($\Xi$), $r_{1,2}+r_{3,2}$ ($\Lambda$),  $r_{2,1}$ (V), and for $\varGamma_{3,2}$ the contributions are $r_{3,2}$ ($\Xi$), $r_{1,2}+r_{3,2}$ ($\Lambda$),  $r_{2,3}$ (V).  Only the occupation-singlet and -triplet equations change their structure, because the occupations explicitly couple to each other. Since the Raman process does not explicitly couple to the occupations, we do not discuss the modifications here. In V-type systems, of course, we need a robust initial occupation of level 1 to efficiently drive the Raman process. In total, we can state that radiative decay will lead to configuration- and device-specific changes of the quantum yield which have to be considered individually for any analysis of experiments. If necessary, further phonon-induced effects can be included\cite{gawarecki2012dephasing,heinze2017polarization}.

\subsection{Quasi-stationary solution}\label{app:Quasistat}
To gain first insight into the fundamental Raman-emission characteristics of the two system classes, we simplify the SDT equations for the limit of weak excitations and stationary electronic occupations ($\fuenf=1$, $\langle P_{2,2}\rangle=\sechs=0$) such that we can neglect the triplet order and the backaction of the cavity field with the system. The essential properties found in this regime also survive in the non-stationary limit. Temporarily, let $N\equiv\eins$ (photon population), $R\equiv\dreienv$ (Raman correlation), $\varPi\equiv\zweiL$ or $\zweiV$ (photon-assisted polarization of $\Lambda$ or V), and $P\equiv\vierLenv$ or $\vierVenv$ ($\Lambda$- or V-polarization induced by the control laser). For $\Lambda$-type systems, this yields
{\allowdisplaybreaks
\begin{align}
 \ddt N&=-\kappa N+\frac{2g}{\hbar}\Im \varPi,\\
 \ddt \varPi&=-\i\left(\frac{\DC}{\hbar}-\i\frac{\kappa+\varGamma_{3,2}}{2}\right)\varPi-\frac{\i\Oenv}{\hbar}R,\notag\\
\ddt R&=-\i\left(\pm\frac{\DL}{\hbar}-\i\frac{\kappa+\varGamma_{3,1}}{2}\right)R-\frac{\i\Oenv}{\hbar}\varPi+ \frac{\i g}{\hbar}P,\notag\\
\ddt P&=-\i\left(\frac{-\DC\pm\DL}{\hbar}-\i\frac{\varGamma_{2,1}}{2}\right)P+\frac{\i\Oenv}{\hbar}+\frac{\i g}{\hbar}R,
\end{align}
}
and for the V-type arrangements
{\allowdisplaybreaks
\begin{align}
 \ddt N&=-\kappa N+\frac{2g}{\hbar}\Im \varPi,\\
 \ddt \varPi&=-\i\left(\frac{\DC}{\hbar}-\i\frac{\kappa+\varGamma_{2,1}}{2}\right)\varPi+\frac{\i\Oenv}{\hbar}R+ \frac{\i g}{\hbar},\notag\\
\ddt R&=-\i\left(\pm\frac{\DL}{\hbar}-\i\frac{\kappa+\varGamma_{3,1}}{2}\right)R+\frac{\i\Oenv}{\hbar}\varPi.\notag\\
\end{align}
}
Note that, in the V-equations, the polarization does not enter in contrast to the $\Lambda$-equations. Furthermore, due to the occupied initial state, we find a constant incoherent source term in the $\varPi$-dynamics of the V-type system in the quasi-stationary limit whereas in the $\Lambda$-class, $\varPi$ can only be initialized via the Raman process driven by the laser polarization $P$. In the cw limit, $\varOmega_\env(t)=\varOmega_0=$ const, we can easily solve both sets of equations and analyze the Raman-correlation equation. For $\Lambda$-type systems, we find
{\allowdisplaybreaks
\begin{align}\label{eq:RLcw}
 |R|^2&=\frac{g^2\varOmega_0^2}{(\DC\mp\DL)^2+\left(\hbar\frac{\varGamma_{2,1}}{2}\right)^2}\left[\left(\mp\DL+\frac{\DC\varOmega_0^2}{\DC^2+\left(\hbar\frac{\kappa+\varGamma_{3,2}}{2}\right)^2}\right.\right.\notag\\
         &\quad-\left.\frac{(\DC\mp\DL)g^2}{(\DC\mp\DL)^2+\left(\hbar\frac{\varGamma_{2,1}}{2}\right)^2}\right)^2+\left(\frac{\hbar\frac{\kappa+\varGamma_{3,2}}{2}\varOmega_0^2}{\DC^2+\left(\hbar\frac{\kappa+\varGamma_{3,2}}{2}\right)^2}\right.\notag\\
       &\quad+\left.\left.\frac{\hbar\frac{\varGamma_{2,1}}{2}g^2}{(\DC\mp\DL)^2+\left(\hbar\frac{\varGamma_{2,1}}{2}\right)^2}+\hbar\frac{\kappa+\varGamma_{3,1}}{2}\right)^2\right]^{-1},
\end{align}
}
and for V-type systems we obtain
{\allowdisplaybreaks
\begin{align}\label{eq:RVcw}
 |R|^2&=\frac{g^2\varOmega_0^2}{\DC^2+\left(\hbar\frac{\kappa+\varGamma_{2,1}}{2}\right)^2}\left[\left(\mp\DL+\frac{\DC\varOmega_0^2}{\DC^2+\left(\hbar\frac{\kappa+\varGamma_{2,1}}{2}\right)^2}\right)^2\right.\notag\\
       &\quad+\left.\left(\frac{\hbar\frac{\kappa+\varGamma_{2,1}}{2}\varOmega_0^2}{\DC^2+\left(\hbar\frac{\kappa+\varGamma_{2,1}}{2}\right)^2}+\hbar\frac{\kappa+\varGamma_{3,1}}{2}\right)^2\right]^{-1}.
\end{align}
{\allowdisplaybreaks
If the $g^2$-shift is negligibly small  in Eq.~\eqref{eq:RLcw} (e.g., if $g$ is small or $\DC$ is significantly greater than $g$), we can identify a common expression for the optimum-laser detuning of both system classes,
\begin{equation}
 \DL^\mathrm{cw}=\pm\frac{\DC\varOmega_0^2}{\DC^2+\left(\hbar\frac{\kappa+\varGamma_{i,j}}{2}\right)^2},
\label{eq:DLoptapp}
\end{equation}
with $(i,j)=(2,1)$ for V- and $(i,j)=(3,2)$ for $\Lambda$-type systems, and $-$ for $\Xi$-  and + for $\Lambda$- and V-configurations. We note that in any system configuration, the laser frequency -- i.e. the length of the blue arrows in Fig. \ref{fig:detunings} -- has to be tuned towards the intermediate state 2 by $|\DL^\mathrm{cw}|$ with respect to the bare resonance condition to achieve optimum Raman-photon emission. We also recognize a laser-induced additional damping of the Raman signal which should be minimized via the choice of $\DC$ and $\varOmega_0$ in comparison with the natural damping $(\kappa+\varGamma_{3,1})/2$. In particular, for a fixed control amplitude, we observe enhanced damping for vanishing cavity detuning.

\bibliography{Bibliography.bib}

\end{document}